\documentclass[sn-standardnature]{sn-jnl}

\usepackage{siunitx}
\usepackage{amssymb}
\usepackage{times}
\usepackage{graphicx}
\usepackage{makecell}
\usepackage{rotating}
\usepackage{multibib} 
\usepackage{anyfontsize}
\newcites{App}{Methods References}

\jyear{2024}%

\usepackage{soul}

\usepackage{xr-hyper}
\begin{document}

\title{\centering{Hydrogen sulfide and metal-enriched atmosphere for a Jupiter-mass exoplanet}}


\author*[1]{\fnm{Guangwei} \sur{Fu}}\email{guangweifu@gmail.com}

\author[2]{Luis Welbanks}
\author[3]{Drake Deming}
\author[4]{Julie Inglis}
\author[5]{Michael Zhang}
\author[6]{Joshua Lothringer}
\author[3]{Jegug Ih}
\author[8]{Julianne I. Moses}
\author[7]{Everett Schlawin}
\author[4]{Heather A. Knutson}
\author[9]{Gregory Henry}
\author[10]{Thomas Greene}
\author[1]{David K. Sing}
\author[3]{Arjun B. Savel}
\author[3]{Eliza M.-R. Kempton}
\author[11]{Dana R. Louie}
\author[2]{Michael Line}
\author[3]{Matt Nixon}

\affil*[1]{Department of Physics and Astronomy, Johns Hopkins University, Baltimore, MD, USA}
\affil[2]{School of Earth and Space Exploration, Arizona State University, Tempe, AZ, USA}
\affil[3]{Department of Astronomy, University of Maryland, College Park, MD, USA}
\affil[4]{Division of Geological and Planetary Sciences, California Institute of Technology, Pasadena, CA, USA}
\affil[5]{Department of Astronomy \& Astrophysics, University of Chicago, Chicago, IL, USA}
\affil[6]{Department of Physics, Utah Valley University, Orem, UT, USA}
\affil[7]{Steward Observatory, University of Arizona, Tucson, AZ, USA}
\affil[8]{Space Science Institute, Boulder, CO, USA}
\affil[9]{Center of Excellence in Information Systems, Tennessee State University, Nashville, TN, USA}
\affil[10]{NASA Ames Research Center, Moffett Field, CA, USA}
\affil[11]{NASA Goddard Space Flight Center, Greenbelt, MD, USA}


\abstract{As the closest transiting hot Jupiter to Earth, HD~189733b has been the benchmark planet for atmospheric characterization \cite{sing_hubble_2011, deming_strong_2006, knutson_multiwavelength_2008}. It has also been the anchor point for much of our theoretical understanding of exoplanet atmospheres from composition \cite{fortney_unified_2008}, chemistry \cite{moses_disequilibrium_2011, tsai_vulcan_2017}, aerosols \cite{line_influence_2016} to atmospheric dynamics \cite{showman_atmospheric_2009}, escape \cite{lampon_modelling_2021} and modeling techniques \cite{zhang_platon_2020, line_systematic_2014}. Prior studies of HD 189733b have detected carbon and oxygen-bearing molecules H$_2$O and CO \cite{birkby_detection_2013, mccullough_water_2014} in the atmosphere. The presence of CO$_2$ and CH$_4$ has been claimed \cite{swain_presence_2008, swain_molecular_2009} but later disputed \cite{birkby_detection_2013, brogi_rotation_2016, crouzet_water_2014}. The inferred metallicity based on these measurements, a key parameter in tracing planet formation locations \cite{mordasini_imprint_2016}, varies from depletion \cite{madhusudhan_h2o_2014, brogi_retrieving_2019} to enhancement \cite{fisher_retrieval_2018, finnerty_atmospheric_2023}, hindered by limited wavelength coverage and precision of the observations. Here we report detections of H$_2$O (13.4 sigma), CO$_2$ (11.2 sigma), CO (5 sigma), and H$_2$S (4.5 sigma) in the transmission spectrum (2.4-5 micron) of HD 189733b. With an equilibrium temperature of $\sim$1200K, H$_2$O, CO, and H$_2$S are the main reservoirs for oxygen, carbon, and sulfur. Based on the measured abundances of these three major volatile elements, we infer an atmospheric metallicity of 3-5 times stellar. The upper limit on the methane abundance at 5 sigma is 0.1 ppm which indicates a low carbon-to-oxygen ratio ($<$0.2), suggesting formation through the accretion of water-rich icy planetesimals. The low oxygen-to-sulfur and carbon-to-sulfur ratios also support the planetesimal accretion formation pathway \cite{crossfield_volatile--sulfur_2023}.}




\maketitle

We observed two transits of HD 189733b in JWST program 1633 using JWST NIRCam grism F444W and F322W2 filters on August 25 and 29th 2022. The first visit with F444W used \texttt{SUBGRISM64} subarray lasting 7877 integrations with 4 BRIGHT1 groups per integration. Each effective integration is 2.4s for a total effective exposure time of 18780.9s and a total exposure duration of 21504.2s ($\sim$6 hrs) including overhead. The second visit with F322W2 used \texttt{SUBGRISM64} subarray lasting 10437 integrations with 3 BRIGHT1 groups per integration. Each effective integration is 1.7s for a total effective exposure time of 17774.7s and a total exposure duration of 21383.1s ($\sim$6 hrs) including overhead.  The transit duration of HD189733 b is $\sim$1.8 hrs and both observations had additional pre-ingress baseline relative to post-egress baseline in anticipating the potential ramp systematics at the beginning of the exposure from NIRCam infrared detectors. \\

\begin{figure*}
\centering
  \includegraphics[width=\textwidth,keepaspectratio]{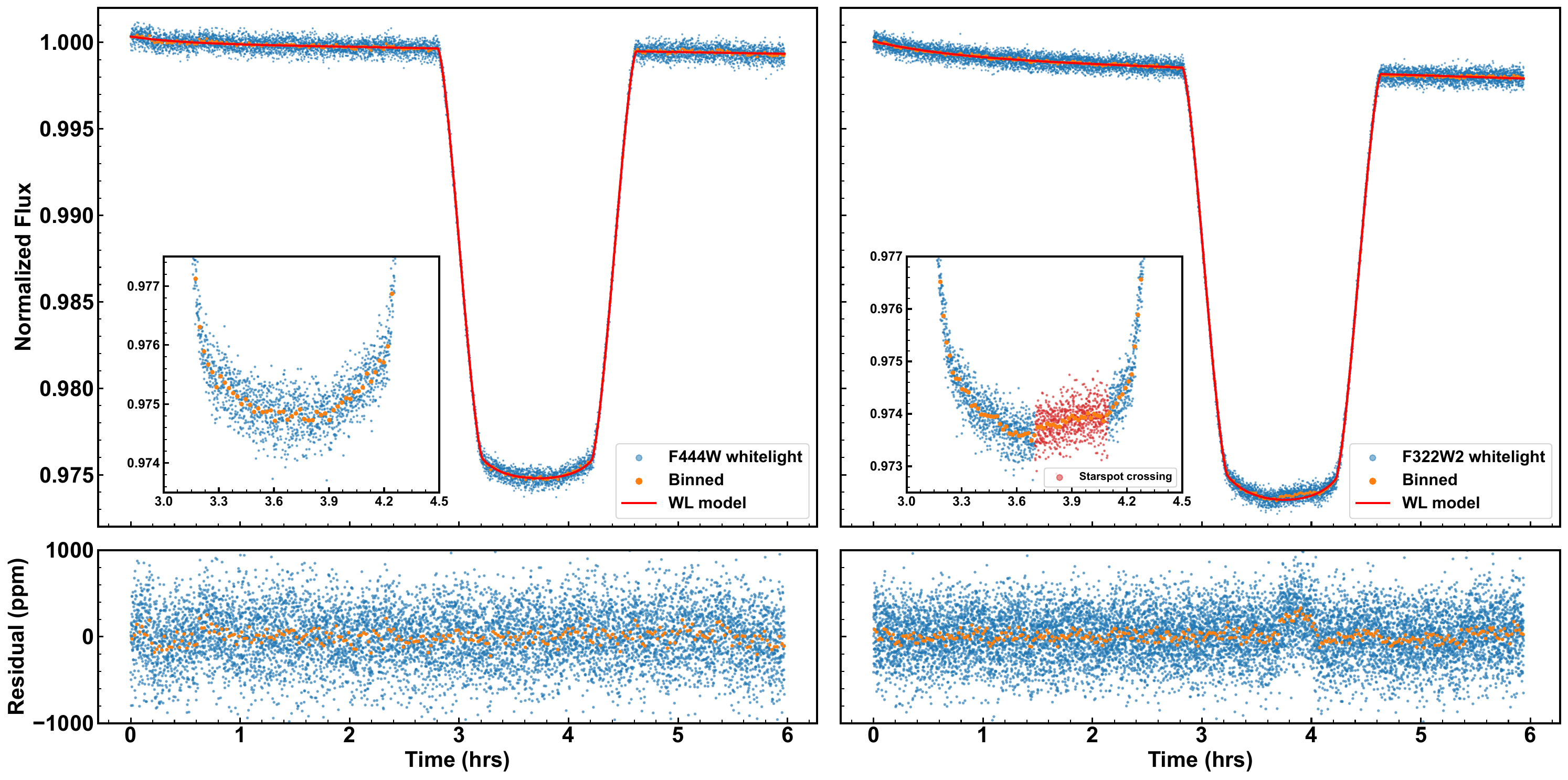}
  \caption{\textbf{The whitelight curve of visit 1 and 2 with F444W (left) and F322W2 (right) respectively.} The F322W2 visit shows a stronger instrumental ramp feature at the beginning of the exposure compared to the F444W visit. There is also a visible star spot crossing event during the F322W2 visit.}
  \label{fig:wl}
\end{figure*}

JWST NIRCam time-series observations (TSO) have exhibited ramp systematics at the beginning of the exposure with amplitudes correlating with the brightness of the star and the number of groups used per integration. This effect was minimal in the WASP-39b observation using NIRCam F322W2 \cite{ahrer_early_2022} since this star is relatively faint (K$\sim$10.2) and as a result, each integration contained 12 groups. For comparison, the ramp was more prominent for the HD 149026b observation \cite{bean_high_2023} which used 5 groups per integration for the F322W2 observation. Our observation of HD 189733b (K$\sim$5.5) has 3 groups and 4 groups per integration for F322W2 and F444W. As a result, we see stronger ramp systematics at the beginning of the light curves (Figure \ref{fig:wl}), especially for F322W2. We performed four (A, B, C, D) independently developed data analyses on the long-wavelength (LW) channel covering 2.4 to 5$\mu m$ (see Methods). Three (A, B, D) analyses used the same JWST Science Calibration Pipeline stage one output and the other one (C) did not. All analyses followed the general steps of first extracting the stellar spectrum from each integration and then they were summed in wavelength to create the white light curve. The best-fit mid-transit time and orbital parameters from the white light are then fixed during the fitting of individual spectroscopic channels created by summing over 20 pixels in wavelength per channel (R$\sim$200). The ramp shape is wavelength-dependent for the F322W2 visit (Extended Data Figure \ref{fig:F322_data_model_residual_each_amp}) as seen in all four independent data reduction analyses. The amplitude of the ramp is the largest at around 3 $\mu m$, and the root cause of this detector behavior is currently unknown. The final transmission spectra from the four reductions are in agreement as shown in Extended Data Figure \ref{fig:compare_reductions}. The transmission spectrum from analysis A is used for the model interpretation. \\

We observed a starspot crossing event during the F332W2 transit that lasted $\sim$30 minutes but no visible starspot crossing during the F444W visit. This is consistent with the two visits being separated by $\sim$4.4 days and a stellar rotational period of $\sim$11.8 days. HD 189733 is an active K1.5V star known for heavy starspot coverage \cite{sing_hubble_2011}. We excluded the starspot crossing integrations (Figure \ref{fig:wl}) in the whitelight fit for the F322W2 visit. The starspot residuals are then fitted with a second-order polynomial to model the feature's shape. We scaled the polynomial starspot feature model from the whitelight residual to fit the starspot feature in each wavelength channel \cite{sing_hubble_2011, fu_water_2022}. We also tested masking out the starspot crossing event in the light curves and achieved good agreement in the resulting transmission spectra (see Methods). \\

\begin{figure*}[h!]
\centering
  \includegraphics[width=0.9\textwidth,keepaspectratio]{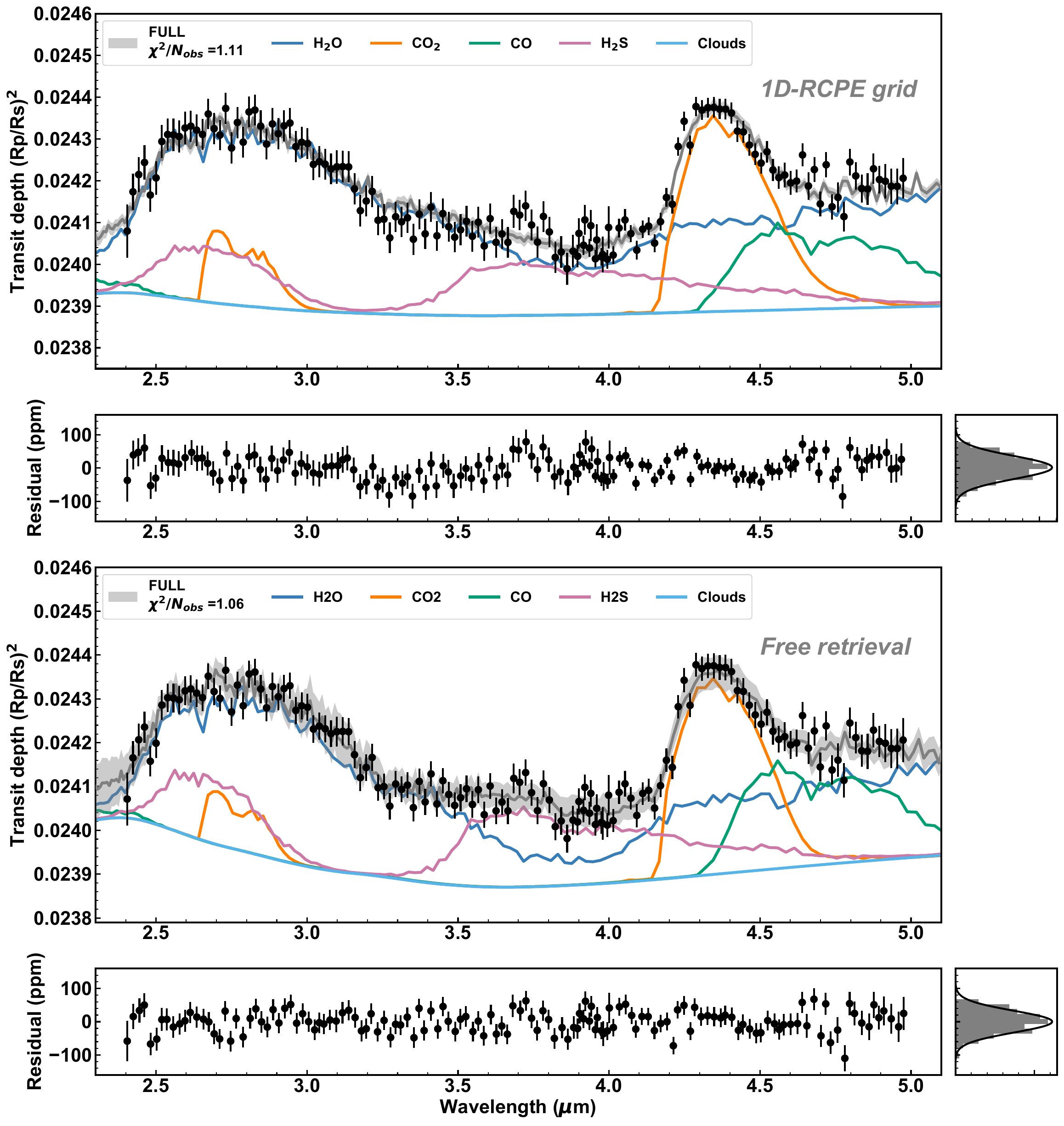}
  \caption{\textbf{The JWST NIRCam transmission spectrum of HD 189733b.} In the top two panels, the spectrum and residuals from analysis A are shown in black points. The best-fit model from the 1-dimensional radiative-convective photo-chemical-equilibrium (1D-RCPE) grid (Tint=500K with disequilibrium chemistry) is shown in grey. Models with contributions only from H$_2$O, CO$_2$, CO, H$_2$S, and clouds are overplotted in corresponding colors. The middle two panels are similar to the top two panels but with the best-fit free retrieval model and the 1$\sigma$ confidence band shown in grey. The free retrieval provides a better fit due to the more flexible model assumptions allowing each molecular abundance to vary independently. N$_{obs}$ is 139 data points with 10 free parameters used in the grid retrieval model ($\chi^2_{\nu}$=1.2) and 20 free parameters used in the free retrieval model ($\chi^2_{\nu}$=1.24). The residuals are both consistent with Gaussian distributions. The median uncertainties of the spectra (36.5ppm and 32.6ppm) are 1.05 and 0.94 times the width of the distributions (34.6 and 36.5ppm) for grid and free retrievals respectively.}
  \label{fig:fig2}
\end{figure*}

Our two NIRCam visits are $\sim$4.4 days apart transiting different stellar disks. Different coverage fractions of unocculted spots will induce offsets between the two visits \cite{rackham_transit_2018}. We expect the wavelength-dependent effects of unocculted spots to be small at these long wavelengths \cite{fu_water_2022}. Based on the ground-based photometric monitor (Supplementary Information Figure \ref{fig:fig3_APT}), we estimate the unocculted starspots to cause the star to fade $\sim$2.7 percent at the first visit and $\sim$1 percent at the second visit relative to when the star is at the brightest. We then applied a dilution correction using equations (4) and (5) from Ref. \cite{sing_hubble_2011} assuming 2.7 and 1 percent dimming and starspot temperature of 4250K measured by STIS \cite{pont_prevalence_2013} to F444W and F322W2 spectra respectively. In addition to stellar heterogeneity, nightside emission from the planet can also pollute the transit spectrum \cite{kipping_nightside_2010}. While we do not expect this effect to induce offsets between visits since the nightside temperature of the planet does not vary strongly in time \cite{komacek_temporal_2019}, the nightside dilution does have a strong wavelength dependency as the flux ratio between the nightside relative to the star increases with wavelength. We applied the nightside correction by adopting formula (4) in Ref. \cite{kipping_nightside_2010} and assumed a blackbody nightside spectrum with a temperature of 1011K based on the Spitzer 8 $\mu$m phase curve measurement \cite{knutson_multiwavelength_2008}. The effects of both nightside dilution and stellar heterogeneity corrections on the transmission spectrum are shown in the Extended Data Figure \ref{fig:correction_plot}. Any additional residuals in the correction from uncertainties in stellar heterogeneity and nightside molecular absorption are likely to be small and will be marginalized to the first order by fitting the offset between two visits within retrievals. \\

\begin{figure*}
\centering
  \includegraphics[width=0.9\textwidth,keepaspectratio]{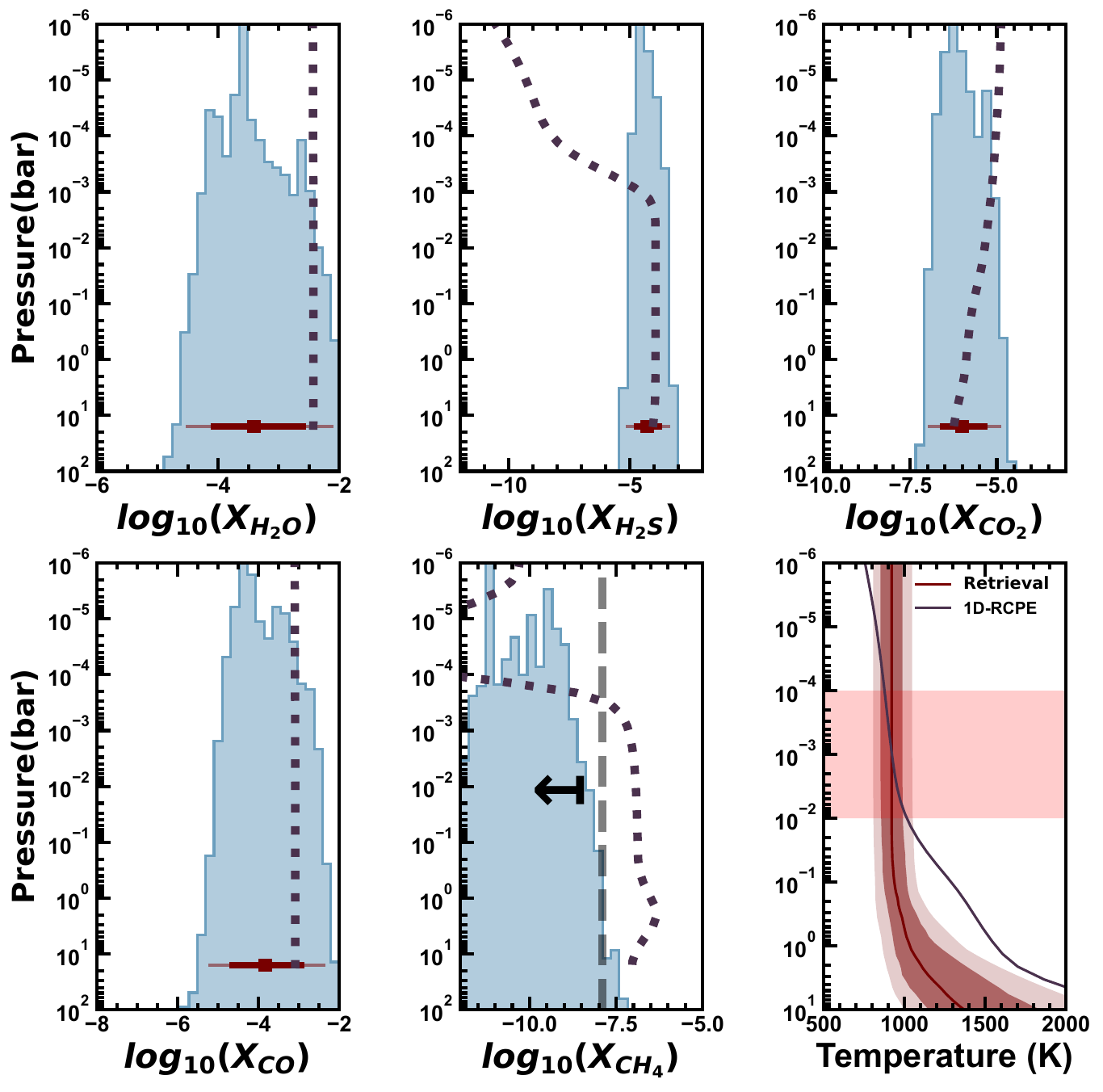}
  \caption{\textbf{The best-fit grid model VMR compared to free retrieval posterior.} The dotted lines represent the volume mixing ratio (VMR) from the best-fit 1D-RCPE grid model as shown in Figure \ref{fig:fig2}. The blue-shaded regions are the free retrieval posterior distribution for each corresponding chemical species. The red horizontal points are the median value of each posterior with the solid and shaded error bars representing 1 and 2 sigma uncertainties. The 99\% upper limit on the abundance of CH$_4$ is shown by the gray dashed line and black arrow. The pink-shaded region in the temperature-pressure (TP) plot indicates the pressure levels probed by the transmission spectrum.}
  \label{fig:fig3}
\end{figure*}

The resulting NIRCam transmission spectrum shown in Figure~\ref{fig:fig2} shows two prominent features at 2.7$\mu$m and 4.3$\mu$m with an increase in transit depth of close to 400 ppm.  A comparison to the cross-sections of the main molecules expected to be present in hot Jupiters at these temperatures \cite{burrows_spectra_2014} suggests that these features are due to H$_2$O and CO$_2$ absorption in the atmosphere of HD~189733b. Other gases such as CO and H$_2$S have significant contributions to the overall continuum level of the spectrum. The signatures of other gases such as CH$_4$, NH$_3$, and HCN are not visible in this NIRCam spectrum.

To interpret these observations and quantify the detections of the chemical species present in the atmosphere of HD~189733b, we took two complementary modeling approaches. They both use the Bayesian inference methodologies known as atmospheric retrievals \cite{madhusudhan_atmospheric_2018} but with different model assumptions. In the first retrieval approach, we use a grid of self-consistent atmospheric models for HD~189733b assuming 1-dimensional radiative-convective photo-chemical-equilibrium (1D-RCPE). The inherent assumption of 1D-RCPE allows for eliminating unlikely chemical and thermal solutions but is limited in its flexibility due to the minimal number of free parameters in the model. In the second retrieval approach, atmospheric models are generated without the 1D-RCPE assumption \cite{line_systematic_2013}, and the abundance of each molecule is allowed to freely vary independently. This methodology ``free of assumptions" is commonly known as a ``free retrieval". The higher degrees of freedom within free retrieval usually result in a better fit to the data but could lead to less physically and chemically motivated combinations of solutions.


\begin{figure*}
\centering
  \includegraphics[width=\textwidth,keepaspectratio]{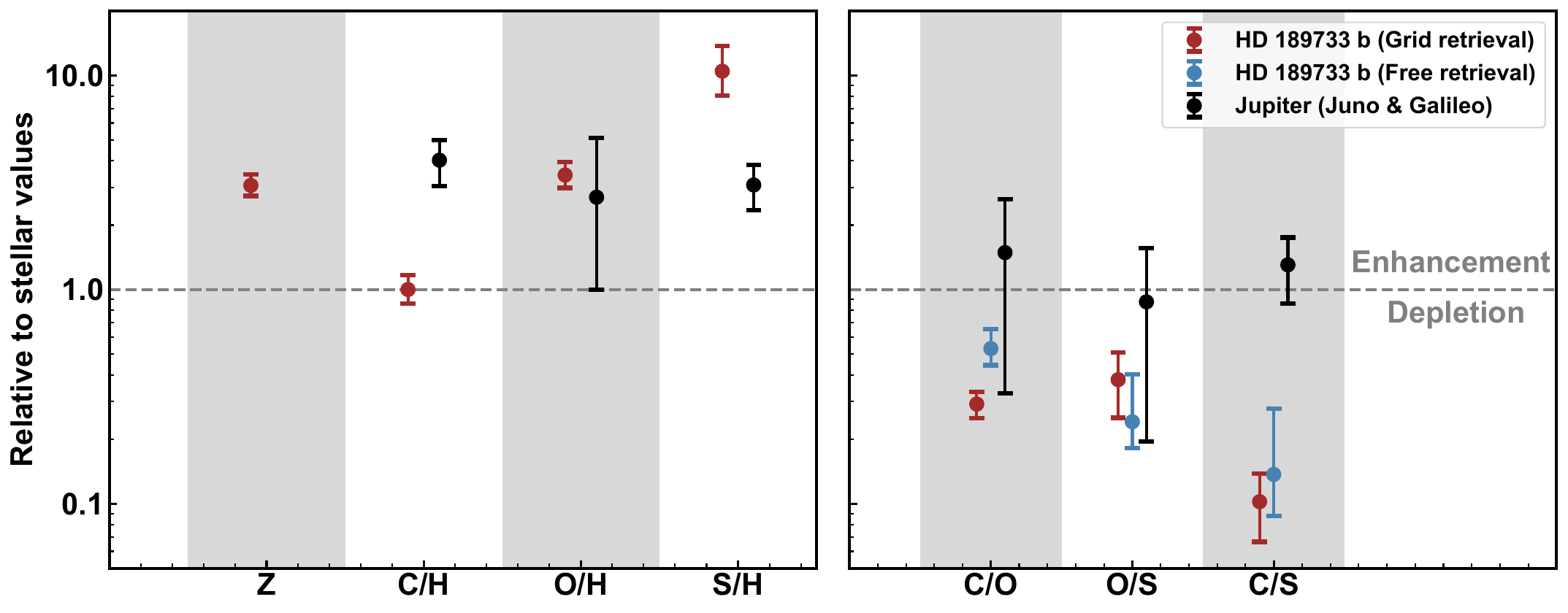}
  \caption{\textbf{Inferred atmospheric elemental abundances of HD 189733b.} The left panel shows the overall metallicity (Z) and individual C/H, O/H, and S/H values relative to the host star. The right panel shows the elemental ratios C/O, O/S, and C/S relative to the host star. The grey dashed line indicates the stellar values, above the line represents metal enhancement, and below represents depletion. The host star HD 189733 has a composition consistent with solar in both C/O and metallicity \cite{brewer_co_2016}.  Measurements of Jupiter from Juno and Galileo probes \cite{atreya_origin_2016, li_water_2020} are overplotted in black for reference. The free retrieval inferred absolute abundances are not informative and are not plotted in the left panel. This is due to the inherent degeneracy between abundances when all elements can freely vary independently. However, the elemental ratios from free retrieval are robust as they are determined directly by the corresponding molecular feature sizes in the spectrum. The grid sulfur values are based on the independent free-fitting H$_2$S scaling factor for the grid retrieval (See methods). The resulting O/S and C/S ratios both agree with the free retrieval to within one sigma. Both methods give sub-stellar C/O ratios with free retrieval producing a higher value due to larger CO abundance.}
  \label{fig:fig4}
\end{figure*}

Our first 1D-RCPE grid retrieval approach finds that the atmosphere of HD~189733b is best-matched by models with metallicities between $\sim3-5\times$ stellar (at $\pm2\sigma$) and a significantly sub-stellar C/O ratio ($<$ 0.2 at 5$\sigma$) (Extended Data Figure \ref{fig:grid_corner_inflate}). We use a Bayesian nested model comparison to quantify the model preference for the spectroscopic contribution of different molecular species. This model comparison results in $>10\sigma$ detections of H$_2$O and CO$_2$ and detections of H$_2$S at $4.5\sigma$ and CO at $5\sigma$. 

Our second free retrieval approach uses CHIMERA \cite{line_systematic_2013}, which computes an atmospheric model considering the absorption due to H$_2$O, CH$_4$,  CO, CO$_2$, NH$_3$, HCN, SO$_2$, C$_2$H$_2$, and H$_2$S assuming constant-with-altitude volume mixing ratio, a non-isothermal pressure-temperature profile following the parameterization of \cite{parmentier_non-grey_2014}, H$_2$-H$_2$ and H$_2$-He collision-induced absorption, and inhomogeneous clouds and hazes. A Bayesian nested model comparison using this methodology results in a similar detection of CO$_2$ at $10.7\sigma$, CO at $4.9\sigma$, H$_2$S at $5.3\sigma$ and a more conservative, yet still conclusive detection of H$_2$O at $9.6\sigma$ (Extended Data Figure \ref{fig:free_corner_inflate}). The stringent 99\% upper limit on the inferred CH$_4$ abundance of 1.3$\times10^{-8}$ is in strong disagreement with 5$\times$10$^{-5}$ CH$_4$ abundance reported in Ref. \cite{swain_presence_2008}. The non-detection of SO$_2$ is consistent with 3-5 times stellar metallicity in the presence of photochemistry. Since the 4$\mu m$ SO$_2$ feature is strongly metallicity \cite{tsai_photochemically_2023} and C/O ratio \cite{polman_h_2023} dependent, the relatively low atmospheric metallicity of HD 189733b compared to WASP-39b ($\geq$10x solar) can explain why we detected the 4$\mu m$ SO$_2$ feature on WASP-39b \cite{powell_sulfur_2024} but not on HD 189733b. 

To ensure the robustness of our model inferences in the presence of underestimated error bars, we performed the retrievals including one error inflation parameter each for F322W and F444W. For the grid retrieval, we retrieved median values of  29.5 and 0 (unconstrained) ppm inflation terms for F322W and F444W respectively. For the free retrieval, we retrieved median values of 20.9 and 14.8 ppm inflation terms for F322W and F444W respectively. The inflated error terms are added in quadrature to the existing error values.

We found a low methane abundance upper limit and a low C/O ratio in the atmosphere of HD189733b. We explored additional mechanisms that could give rise to the observed low methane abundance such as photochemistry and vertical mixing with high internal temperature assumption (Tint=500K) in our disequilibrium chemistry grid calculation (See Methods). However, the low C/O solution is preferred in the presence of these processes as photochemistry based on the assumed UV flux model \cite{moses_disequilibrium_2011} does not sufficiently deplete CH$_4$ and vertical mixing enhances, rather than depletes CH$_4$, due to higher abundance deeper down in the atmosphere. While we did not include the effect of horizontal transport in this study, General Circulation Model (GCM) predictions predict enhanced CH$_4$ abundances rather than the observed depletion \cite{lee_mini-chemical_2023}. The enhanced super-stellar metallicity is primarily driven by the prominent CO$_2$ and H$_2$O absorption feature. As both molecules are insensitive to the effects of disequilibrium processes \cite{moses_disequilibrium_2011}, the relative feature strength of the two is a robust tracer for atmospheric metallicity. The combination of inferred sub-stellar C/O and super-stellar metallicity of HD 189733b (Figure \ref{fig:fig4}) supports a formation history of accreting water-rich icy planetesimals \cite{oberg_effects_2011, mordasini_imprint_2016, madhusudhan_co_2012}.



\bibliographystyle{sn-standardnature}

\bibliography{references.bib} 

\begin{thebibliography}{10}
\expandafter\ifx\csname url\endcsname\relax
  \def\url#1{\burl{#1}}\fi
\expandafter\ifx\csname urlprefix\endcsname\relax\def\urlprefix{URL }\fi
\providecommand{\bibinfo}[2]{#2}
\providecommand{\eprint}[2][]{\url{#2}}
\providecommand{\doi}[1]{\url{https://doi.org/#1}}
\bibcommenthead

\bibitem{sing_hubble_2011}
\bibinfo{author}{Sing, D.~K.} \emph{et~al.}
\newblock \bibinfo{title}{Hubble {Space} {Telescope} {Transmission} {Spectroscopy} of the {Exoplanet} {HD} 189733b: {High}-altitude atmospheric haze in the optical and near-{UV} with {STIS}}.
\newblock \emph{\bibinfo{journal}{Monthly Notices of the Royal Astronomical Society}} \textbf{\bibinfo{volume}{416}}~(2), \bibinfo{pages}{1443--1455} (\bibinfo{year}{2011}).
\newblock \bibinfo{note}{ArXiv: 1103.0026}.

\bibitem{deming_strong_2006}
\bibinfo{author}{Deming, D.}, \bibinfo{author}{Harrington, J.}, \bibinfo{author}{Seager, S.} \& \bibinfo{author}{Richardson, L.~J.}
\newblock \bibinfo{title}{Strong {Infrared} {Emission} from the {Extrasolar} {Planet} {HD} 189733b}.
\newblock \emph{\bibinfo{journal}{The Astrophysical Journal}} \textbf{\bibinfo{volume}{644}}~(1), \bibinfo{pages}{560--564} (\bibinfo{year}{2006}).

\bibitem{knutson_multiwavelength_2008}
\bibinfo{author}{Knutson, H.~A.} \emph{et~al.}
\newblock \bibinfo{title}{{MULTIWAVELENGTH} {CONSTRAINTS} {ON} {THE} {DAY}–{NIGHT} {CIRCULATION} {PATTERNS} {OF} {HD} 189733b}.
\newblock \emph{\bibinfo{journal}{The Astrophysical Journal}} \textbf{\bibinfo{volume}{690}}~(1), \bibinfo{pages}{822} (\bibinfo{year}{2008}).
\newblock \bibinfo{note}{Publisher: The American Astronomical Society}.

\bibitem{fortney_unified_2008}
\bibinfo{author}{Fortney, J.}, \bibinfo{author}{Lodders, K.}, \bibinfo{author}{Marley, M.} \& \bibinfo{author}{Freedman, R.}
\newblock \bibinfo{title}{A {Unified} {Theory} for the {Atmospheres} of the {Hot} and {Very} {Hot} {Jupiters}: {Two} {Classes} of {Irradiated} {Atmospheres}}.
\newblock \emph{\bibinfo{journal}{The Astrophysical Journal}} \textbf{\bibinfo{volume}{678}}~(2), \bibinfo{pages}{1419--1435} (\bibinfo{year}{2008}).

\bibitem{moses_disequilibrium_2011}
\bibinfo{author}{Moses, J.~I.} \emph{et~al.}
\newblock \bibinfo{title}{{DISEQUILIBRIUM} {CARBON}, {OXYGEN}, {AND} {NITROGEN} {CHEMISTRY} {IN} {THE} {ATMOSPHERES} {OF} {HD} 189733b {AND} {HD} 209458b}.
\newblock \emph{\bibinfo{journal}{The Astrophysical Journal}} \textbf{\bibinfo{volume}{737}}~(1), \bibinfo{pages}{15} (\bibinfo{year}{2011}).

\bibitem{tsai_vulcan_2017}
\bibinfo{author}{Tsai, S.-M.} \emph{et~al.}
\newblock \bibinfo{title}{{VULCAN} : {An} {Open}-source, {Validated} {Chemical} {Kinetics} {Python} {Code} for {Exoplanetary} {Atmospheres}}.
\newblock \emph{\bibinfo{journal}{The Astrophysical Journal Supplement Series}} \textbf{\bibinfo{volume}{228}}~(2), \bibinfo{pages}{20} (\bibinfo{year}{2017}).

\bibitem{line_influence_2016}
\bibinfo{author}{Line, M.~R.} \& \bibinfo{author}{Parmentier, V.}
\newblock \bibinfo{title}{The {Influence} of {Non}-{Uniform} {Cloud} {Cover} on {Transit} {Transmission} {Spectra}}.
\newblock \emph{\bibinfo{journal}{The Astrophysical Journal}} \textbf{\bibinfo{volume}{820}}~(1), \bibinfo{pages}{78} (\bibinfo{year}{2016}).
\newblock \bibinfo{note}{ArXiv: 1511.09443}.

\bibitem{showman_atmospheric_2009}
\bibinfo{author}{Showman, A.~P.} \emph{et~al.}
\newblock \bibinfo{title}{Atmospheric circulation of hot {Jupiters}: coupled radiative-dynamical general circulation model simulations of {HD} 189733b and {HD} 209458b}.
\newblock \emph{\bibinfo{journal}{The Astrophysical Journal}} \textbf{\bibinfo{volume}{699}}~(1), \bibinfo{pages}{564--584} (\bibinfo{year}{2009}).

\bibitem{lampon_modelling_2021}
\bibinfo{author}{Lampón, M.} \emph{et~al.}
\newblock \bibinfo{title}{Modelling the {He} {I} triplet absorption at 10 830 Å in the atmospheres of {HD} 189733 b and {GJ} 3470 b}.
\newblock \emph{\bibinfo{journal}{Astronomy \& Astrophysics}} \textbf{\bibinfo{volume}{647}}, \bibinfo{pages}{A129} (\bibinfo{year}{2021}).

\bibitem{zhang_platon_2020}
\bibinfo{author}{Zhang, M.} \emph{et~al.}
\newblock \bibinfo{title}{{PLATON} {II}: {New} {Capabilities} {And} {A} {Comprehensive} {Retrieval} on {HD} 189733b {Transit} and {Eclipse} {Data}}.
\newblock \emph{\bibinfo{journal}{The Astrophysical Journal}} \textbf{\bibinfo{volume}{899}}~(1), \bibinfo{pages}{27} (\bibinfo{year}{2020}).
\newblock \bibinfo{note}{ArXiv: 2004.09513}.

\bibitem{line_systematic_2014}
\bibinfo{author}{Line, M.~R.}, \bibinfo{author}{Knutson, H.}, \bibinfo{author}{Wolf, A.~S.} \& \bibinfo{author}{Yung, Y.~L.}
\newblock \bibinfo{title}{A systematic retrieval analysis of secondary eclipse spectra. {II}. a uniform analysis of nine planets and their {C} to {O} ratios}.
\newblock \emph{\bibinfo{journal}{The Astrophysical Journal}} \textbf{\bibinfo{volume}{783}}~(2), \bibinfo{pages}{70} (\bibinfo{year}{2014}).

\bibitem{birkby_detection_2013}
\bibinfo{author}{Birkby, J.~L.} \emph{et~al.}
\newblock \bibinfo{title}{Detection of water absorption in the day side atmosphere of {HD} 189733 b using ground-based high-resolution spectroscopy at 3.2 $\mu$m}.
\newblock \emph{\bibinfo{journal}{Monthly Notices of the Royal Astronomical Society: Letters}} \textbf{\bibinfo{volume}{436}}~(1), \bibinfo{pages}{L35--L39} (\bibinfo{year}{2013}).

\bibitem{mccullough_water_2014}
\bibinfo{author}{McCullough, P.~R.}, \bibinfo{author}{Crouzet, N.}, \bibinfo{author}{Deming, D.} \& \bibinfo{author}{Madhusudhan, N.}
\newblock \bibinfo{title}{Water vapor in the spectrum of the extrasolar planet {HD} 189733b. {I}. {The} transit}.
\newblock \emph{\bibinfo{journal}{The Astrophysical Journal}} \bibinfo{pages}{11} (\bibinfo{year}{2014}).

\bibitem{swain_presence_2008}
\bibinfo{author}{Swain, M.~R.}, \bibinfo{author}{Vasisht, G.} \& \bibinfo{author}{Tinetti, G.}
\newblock \bibinfo{title}{The presence of methane in the atmosphere of an extrasolar planet}.
\newblock \emph{\bibinfo{journal}{Nature}} \textbf{\bibinfo{volume}{452}}~(7185), \bibinfo{pages}{329--331} (\bibinfo{year}{2008}).
\newblock \bibinfo{note}{Number: 7185 Publisher: Nature Publishing Group}.

\bibitem{swain_molecular_2009}
\bibinfo{author}{Swain, M.~R.} \emph{et~al.}
\newblock \bibinfo{title}{{MOLECULAR} {SIGNATURES} {IN} {THE} {NEAR}-{INFRARED} {DAYSIDE} {SPECTRUM} {OF} {HD} 189733b}.
\newblock \emph{\bibinfo{journal}{The Astrophysical Journal}} \textbf{\bibinfo{volume}{690}}~(2), \bibinfo{pages}{L114--L117} (\bibinfo{year}{2009}).

\bibitem{brogi_rotation_2016}
\bibinfo{author}{Brogi, M.} \emph{et~al.}
\newblock \bibinfo{title}{{ROTATION} {AND} {WINDS} {OF} {EXOPLANET} {HD} 189733 b {MEASURED} {WITH} {HIGH}-{DISPERSION} {TRANSMISSION} {SPECTROSCOPY}}.
\newblock \emph{\bibinfo{journal}{The Astrophysical Journal}} \textbf{\bibinfo{volume}{817}}~(2), \bibinfo{pages}{106} (\bibinfo{year}{2016}).

\bibitem{crouzet_water_2014}
\bibinfo{author}{Crouzet, N.}, \bibinfo{author}{McCullough, P.~R.}, \bibinfo{author}{Deming, D.} \& \bibinfo{author}{Madhusudhan, N.}
\newblock \bibinfo{title}{Water vapor in the spectrum of the extrasolar planet {HD} 189733b: 2. {The} eclipse}.
\newblock \emph{\bibinfo{journal}{The Astrophysical Journal}} \textbf{\bibinfo{volume}{795}}~(2), \bibinfo{pages}{166} (\bibinfo{year}{2014}).
\newblock \bibinfo{note}{ArXiv:1409.4000 [astro-ph]}.

\bibitem{mordasini_imprint_2016}
\bibinfo{author}{Mordasini, C.}, \bibinfo{author}{van Boekel, R.}, \bibinfo{author}{Mollière, P.}, \bibinfo{author}{Henning, T.} \& \bibinfo{author}{Benneke, B.}
\newblock \bibinfo{title}{The imprint of exoplanet formation history on observable present-day spectra of hot {Jupiters}}.
\newblock \emph{\bibinfo{journal}{The Astrophysical Journal}} \textbf{\bibinfo{volume}{832}}~(1), \bibinfo{pages}{41} (\bibinfo{year}{2016}).
\newblock \bibinfo{note}{ArXiv:1609.03019 [astro-ph]}.

\bibitem{madhusudhan_h2o_2014}
\bibinfo{author}{Madhusudhan, N.}, \bibinfo{author}{Crouzet, N.}, \bibinfo{author}{McCullough, P.~R.}, \bibinfo{author}{Deming, D.} \& \bibinfo{author}{Hedges, C.}
\newblock \bibinfo{title}{H$_{\textrm{2}}${O} abundances in the atmospheres of three hot {Jupiters}}.
\newblock \emph{\bibinfo{journal}{The Astrophysical Journal}} \textbf{\bibinfo{volume}{791}}~(1), \bibinfo{pages}{L9} (\bibinfo{year}{2014}).

\bibitem{brogi_retrieving_2019}
\bibinfo{author}{Brogi, M.} \& \bibinfo{author}{Line, M.~R.}
\newblock \bibinfo{title}{Retrieving {Temperatures} and {Abundances} of {Exoplanet} {Atmospheres} with {High}-resolution {Cross}-correlation {Spectroscopy}}.
\newblock \emph{\bibinfo{journal}{The Astronomical Journal}} \textbf{\bibinfo{volume}{157}}~(3), \bibinfo{pages}{114} (\bibinfo{year}{2019}).

\bibitem{fisher_retrieval_2018}
\bibinfo{author}{Fisher, C.} \& \bibinfo{author}{Heng, K.}
\newblock \bibinfo{title}{Retrieval analysis of 38 {WFC3} transmission spectra and resolution of the normalization degeneracy}.
\newblock \emph{\bibinfo{journal}{Monthly Notices of the Royal Astronomical Society}} \textbf{\bibinfo{volume}{481}}~(4), \bibinfo{pages}{4698--4727} (\bibinfo{year}{2018}).

\bibitem{finnerty_atmospheric_2023}
\bibinfo{author}{Finnerty, L.} \emph{et~al.}
\newblock \bibinfo{title}{Atmospheric metallicity and {C}/{O} of {HD} 189733 b from high-resolution spectroscopy} (\bibinfo{year}{2023}).
\newblock \urlprefix\url{http://arxiv.org/abs/2312.00141}.
\newblock \bibinfo{note}{ArXiv:2312.00141 [astro-ph]}.

\bibitem{crossfield_volatile--sulfur_2023}
\bibinfo{author}{Crossfield, I. J.~M.}
\newblock \bibinfo{title}{Volatile-to-sulfur {Ratios} {Can} {Recover} a {Gas} {Giant}’s {Accretion} {History}}.
\newblock \emph{\bibinfo{journal}{The Astrophysical Journal Letters}} \textbf{\bibinfo{volume}{952}}~(1), \bibinfo{pages}{L18} (\bibinfo{year}{2023}).

\bibitem{ahrer_early_2022}
\bibinfo{author}{Ahrer, E.-M.} \emph{et~al.}
\newblock \bibinfo{title}{Early {Release} {Science} of the exoplanet {WASP}-39b with {JWST} {NIRCam}} (\bibinfo{year}{2022}).
\newblock \urlprefix\url{http://arxiv.org/abs/2211.10489}.
\newblock \bibinfo{note}{ArXiv:2211.10489 [astro-ph]}.

\bibitem{bean_high_2023}
\bibinfo{author}{Bean, J.~L.} \emph{et~al.}
\newblock \bibinfo{title}{High atmospheric metal enrichment for a {Saturn}-mass planet}.
\newblock \emph{\bibinfo{journal}{Nature}} \bibinfo{pages}{1--2} (\bibinfo{year}{2023}).
\newblock \bibinfo{note}{Publisher: Nature Publishing Group}.

\bibitem{fu_water_2022}
\bibinfo{author}{Fu, G.} \emph{et~al.}
\newblock \bibinfo{title}{Water and an {Escaping} {Helium} {Tail} {Detected} in the {Hazy} and {Methane}-depleted {Atmosphere} of {HAT}-{P}-18b from {JWST} {NIRISS}/{SOSS}}.
\newblock \emph{\bibinfo{journal}{The Astrophysical Journal Letters}} \textbf{\bibinfo{volume}{940}}~(2), \bibinfo{pages}{L35} (\bibinfo{year}{2022}).

\bibitem{rackham_transit_2018}
\bibinfo{author}{Rackham, B.~V.}, \bibinfo{author}{Apai, D.} \& \bibinfo{author}{Giampapa, M.~S.}
\newblock \bibinfo{title}{The {Transit} {Light} {Source} {Effect}: {False} {Spectral} {Features} and {Incorrect} {Densities} for {M}-dwarf {Transiting} {Planets}}.
\newblock \emph{\bibinfo{journal}{The Astrophysical Journal}} \textbf{\bibinfo{volume}{853}}~(2), \bibinfo{pages}{122} (\bibinfo{year}{2018}).
\newblock \bibinfo{note}{Publisher: The American Astronomical Society}.

\bibitem{pont_prevalence_2013}
\bibinfo{author}{Pont, F.} \emph{et~al.}
\newblock \bibinfo{title}{The prevalence of dust on the exoplanet {HD} 189733b from {Hubble} and {Spitzer} observations}.
\newblock \emph{\bibinfo{journal}{Monthly Notices of the Royal Astronomical Society}} \textbf{\bibinfo{volume}{432}}~(4), \bibinfo{pages}{2917--2944} (\bibinfo{year}{2013}).

\bibitem{kipping_nightside_2010}
\bibinfo{author}{Kipping, D.~M.} \& \bibinfo{author}{Tinetti, G.}
\newblock \bibinfo{title}{Nightside pollution of exoplanet transit depths: {Nightside} pollution of exoplanet transits}.
\newblock \emph{\bibinfo{journal}{Monthly Notices of the Royal Astronomical Society}} \textbf{\bibinfo{volume}{407}}~(4), \bibinfo{pages}{2589--2598} (\bibinfo{year}{2010}).

\bibitem{komacek_temporal_2019}
\bibinfo{author}{Komacek, T.~D.} \& \bibinfo{author}{Showman, A.~P.}
\newblock \bibinfo{title}{Temporal {Variability} in {Hot} {Jupiter} {Atmospheres}}.
\newblock \emph{\bibinfo{journal}{The Astrophysical Journal}} \textbf{\bibinfo{volume}{888}}~(1), \bibinfo{pages}{2} (\bibinfo{year}{2019}).
\newblock \bibinfo{note}{Publisher: The American Astronomical Society}.

\bibitem{burrows_spectra_2014}
\bibinfo{author}{Burrows, A.}
\newblock \bibinfo{title}{Spectra as {Windows} into {Exoplanet} {Atmospheres}}.
\newblock \emph{\bibinfo{journal}{Proceedings of the National Academy of Sciences}} \textbf{\bibinfo{volume}{111}}~(35), \bibinfo{pages}{12601--12609} (\bibinfo{year}{2014}).
\newblock \bibinfo{note}{ArXiv: 1312.2009}.

\bibitem{madhusudhan_atmospheric_2018}
\bibinfo{author}{Madhusudhan, N.}
\newblock \bibinfo{title}{ in \textit{Atmospheric {Retrieval} of {Exoplanets}}}  \bibinfo{pages}{2153--2182} (\bibinfo{year}{2018}).
\newblock \urlprefix\url{http://arxiv.org/abs/1808.04824}.
\newblock \bibinfo{note}{ArXiv:1808.04824 [astro-ph]}.

\bibitem{line_systematic_2013}
\bibinfo{author}{Line, M.~R.} \emph{et~al.}
\newblock \bibinfo{title}{A {Systematic} {Retrieval} {Analysis} of {Secondary} {Eclipse} {Spectra} {I}: {A} {Comparison} of {Atmospheric} {Retrieval} {Techniques}}.
\newblock \emph{\bibinfo{journal}{The Astrophysical Journal}} \textbf{\bibinfo{volume}{775}}~(2), \bibinfo{pages}{137} (\bibinfo{year}{2013}).
\newblock \bibinfo{note}{ArXiv: 1304.5561}.

\bibitem{brewer_co_2016}
\bibinfo{author}{Brewer, J.~M.} \& \bibinfo{author}{Fischer, D.~A.}
\newblock \bibinfo{title}{C/{O} {AND} {Mg}/{Si} {RATIOS} {OF} {STARS} {IN} {THE} {SOLAR} {NEIGHBORHOOD}}.
\newblock \emph{\bibinfo{journal}{The Astrophysical Journal}} \textbf{\bibinfo{volume}{831}}~(1), \bibinfo{pages}{20} (\bibinfo{year}{2016}).
\newblock \bibinfo{note}{Publisher: The American Astronomical Society}.

\bibitem{atreya_origin_2016}
\bibinfo{author}{Atreya, S.~K.} \emph{et~al.}
\newblock \bibinfo{title}{The {Origin} and {Evolution} of {Saturn}, with {Exoplanet} {Perspective}} (\bibinfo{year}{2016}).
\newblock \urlprefix\url{http://arxiv.org/abs/1606.04510}.
\newblock \bibinfo{note}{ArXiv:1606.04510 [astro-ph]}.

\bibitem{li_water_2020}
\bibinfo{author}{Li, C.} \emph{et~al.}
\newblock \bibinfo{title}{The water abundance in {Jupiter}’s equatorial zone}.
\newblock \emph{\bibinfo{journal}{Nature Astronomy}} \textbf{\bibinfo{volume}{4}}~(6), \bibinfo{pages}{609--616} (\bibinfo{year}{2020}).
\newblock \bibinfo{note}{Number: 6 Publisher: Nature Publishing Group}.

\bibitem{parmentier_non-grey_2014}
\bibinfo{author}{Parmentier, V.} \& \bibinfo{author}{Guillot, T.}
\newblock \bibinfo{title}{A non-grey analytical model for irradiated atmospheres: {I}. {Derivation}}.
\newblock \emph{\bibinfo{journal}{Astronomy \& Astrophysics}} \textbf{\bibinfo{volume}{562}}, \bibinfo{pages}{A133} (\bibinfo{year}{2014}).

\bibitem{tsai_photochemically_2023}
\bibinfo{author}{Tsai, S.-M.} \emph{et~al.}
\newblock \bibinfo{title}{Photochemically produced {SO2} in the atmosphere of {WASP}-39b}.
\newblock \emph{\bibinfo{journal}{Nature}} \textbf{\bibinfo{volume}{617}}~(7961), \bibinfo{pages}{483--487} (\bibinfo{year}{2023}).
\newblock \bibinfo{note}{Number: 7961 Publisher: Nature Publishing Group}.

\bibitem{polman_h_2023}
\bibinfo{author}{Polman, J.}, \bibinfo{author}{Waters, L. B. F.~M.}, \bibinfo{author}{Min, M.}, \bibinfo{author}{Miguel, Y.} \& \bibinfo{author}{Khorshid, N.}
\newblock \bibinfo{title}{H $_{\textrm{2}}$ {S} and {SO} $_{\textrm{2}}$ detectability in hot {Jupiters}: {Sulphur} species as indicators of metallicity and {C}/{O} ratio}.
\newblock \emph{\bibinfo{journal}{Astronomy \& Astrophysics}} \textbf{\bibinfo{volume}{670}}, \bibinfo{pages}{A161} (\bibinfo{year}{2023}).

\bibitem{powell_sulfur_2024}
\bibinfo{author}{Powell, D.} \emph{et~al.}
\newblock \bibinfo{title}{Sulfur dioxide in the mid-infrared transmission spectrum of {WASP}-39b}.
\newblock \emph{\bibinfo{journal}{Nature}} \textbf{\bibinfo{volume}{626}}~(8001), \bibinfo{pages}{979--983} (\bibinfo{year}{2024}).
\newblock \bibinfo{note}{Publisher: Nature Publishing Group}.

\bibitem{lee_mini-chemical_2023}
\bibinfo{author}{Lee, E. K.~H.}, \bibinfo{author}{Tsai, S.-M.}, \bibinfo{author}{Hammond, M.} \& \bibinfo{author}{Tan, X.}
\newblock \bibinfo{title}{A mini-chemical scheme with net reactions for {3D} general circulation models - {II}. {3D} thermochemical modelling of {WASP}-39b and {HD} 189733b}.
\newblock \emph{\bibinfo{journal}{Astronomy \& Astrophysics}} \textbf{\bibinfo{volume}{672}}, \bibinfo{pages}{A110} (\bibinfo{year}{2023}).
\newblock \bibinfo{note}{Publisher: EDP Sciences}.

\bibitem{oberg_effects_2011}
\bibinfo{author}{Oberg, K.~I.}, \bibinfo{author}{Murray-Clay, R.} \& \bibinfo{author}{Bergin, E.~A.}
\newblock \bibinfo{title}{The effects of snowlines on {C}/{O} in planetary atmospheres}.
\newblock \emph{\bibinfo{journal}{The Astrophysical Journal}} \textbf{\bibinfo{volume}{743}}~(1), \bibinfo{pages}{L16} (\bibinfo{year}{2011}).

\bibitem{madhusudhan_co_2012}
\bibinfo{author}{Madhusudhan, N.}
\newblock \bibinfo{title}{C/{O} ratio as a dimension for characterizing exoplanetary atmospheres}.
\newblock \emph{\bibinfo{journal}{The Astrophysical Journal}} \textbf{\bibinfo{volume}{758}}~(1), \bibinfo{pages}{36} (\bibinfo{year}{2012}).

\bibitem{schlawin_jwst_2020}
\bibinfo{author}{Schlawin, E.} \emph{et~al.}
\newblock \bibinfo{title}{{JWST} {Noise} {Floor}. {I}. {Random} {Error} {Sources} in {JWST} {NIRCam} {Time} {Series}}.
\newblock \emph{\bibinfo{journal}{The Astronomical Journal}} \textbf{\bibinfo{volume}{160}}~(5), \bibinfo{pages}{231} (\bibinfo{year}{2020}).

\bibitem{kreidberg_batman_2015}
\bibinfo{author}{Kreidberg, L.}
\newblock \bibinfo{title}{batman: {BAsic} {Transit} {Model} {cAlculatioN} in {Python}}.
\newblock \emph{\bibinfo{journal}{Publications of the Astronomical Society of the Pacific}} \textbf{\bibinfo{volume}{127}}~(957), \bibinfo{pages}{1161} (\bibinfo{year}{2015}).

\bibitem{foreman-mackey_emcee_2013}
\bibinfo{author}{Foreman-Mackey, D.}, \bibinfo{author}{Hogg, D.~W.}, \bibinfo{author}{Lang, D.} \& \bibinfo{author}{Goodman, J.}
\newblock \bibinfo{title}{emcee: {The} {MCMC} {Hammer}}.
\newblock \emph{\bibinfo{journal}{Publications of the Astronomical Society of the Pacific}} \textbf{\bibinfo{volume}{125}}~(925), \bibinfo{pages}{306--312} (\bibinfo{year}{2013}).
\newblock \bibinfo{note}{ArXiv: 1202.3665}.

\bibitem{magic_stagger-grid_2015}
\bibinfo{author}{Magic, Z.}, \bibinfo{author}{Chiavassa, A.}, \bibinfo{author}{Collet, R.} \& \bibinfo{author}{Asplund, M.}
\newblock \bibinfo{title}{The {Stagger}-grid: {A} grid of {3D} stellar atmosphere models: {IV}. {Limb} darkening coefficients}.
\newblock \emph{\bibinfo{journal}{Astronomy \& Astrophysics}} \textbf{\bibinfo{volume}{573}}, \bibinfo{pages}{A90} (\bibinfo{year}{2015}).

\bibitem{grant_exo-ticexotic-ld_2022}
\bibinfo{author}{Grant, D.} \& \bibinfo{author}{Wakeford, H.~R., Hannah}.
\newblock \bibinfo{title}{Exo-{TiC}/{ExoTiC}-{LD}: {ExoTiC}-{LD} v3.0.0} (\bibinfo{year}{2022}).
\newblock \urlprefix\url{https://doi.org/10.5281/zenodo.7437681}.

\bibitem{ahrer_early_2022}
\bibinfo{author}{Ahrer, E.-M.} \emph{et~al.}
\newblock \bibinfo{title}{Early {Release} {Science} of the exoplanet {WASP}-39b with {JWST} {NIRCam}} (\bibinfo{year}{2022}).
\newblock \urlprefix\url{http://arxiv.org/abs/2211.10489}.
\newblock \bibinfo{note}{ArXiv:2211.10489 [astro-ph]}.

\bibitem{agol_climate_2010}
\bibinfo{author}{Agol, E.} \emph{et~al.}
\newblock \bibinfo{title}{{THE} {CLIMATE} {OF} {HD} 189733b {FROM} {FOURTEEN} {TRANSITS} {AND} {ECLIPSES} {MEASURED} {BY} \textit{{SPITZER}}}.
\newblock \emph{\bibinfo{journal}{The Astrophysical Journal}} \textbf{\bibinfo{volume}{721}}~(2), \bibinfo{pages}{1861--1877} (\bibinfo{year}{2010}).

\bibitem{sing_hubble_2011}
\bibinfo{author}{Sing, D.~K.} \emph{et~al.}
\newblock \bibinfo{title}{Hubble {Space} {Telescope} {Transmission} {Spectroscopy} of the {Exoplanet} {HD} 189733b: {High}-altitude atmospheric haze in the optical and near-{UV} with {STIS}}.
\newblock \emph{\bibinfo{journal}{Monthly Notices of the Royal Astronomical Society}} \textbf{\bibinfo{volume}{416}}~(2), \bibinfo{pages}{1443--1455} (\bibinfo{year}{2011}).
\newblock \bibinfo{note}{ArXiv: 1103.0026}.

\bibitem{fu_water_2022}
\bibinfo{author}{Fu, G.} \emph{et~al.}
\newblock \bibinfo{title}{Water and an {Escaping} {Helium} {Tail} {Detected} in the {Hazy} and {Methane}-depleted {Atmosphere} of {HAT}-{P}-18b from {JWST} {NIRISS}/{SOSS}}.
\newblock \emph{\bibinfo{journal}{The Astrophysical Journal Letters}} \textbf{\bibinfo{volume}{940}}~(2), \bibinfo{pages}{L35} (\bibinfo{year}{2022}).

\bibitem{arcangeli_new_2021}
\bibinfo{author}{Arcangeli, J.}, \bibinfo{author}{Désert, J.-M.}, \bibinfo{author}{Parmentier, V.}, \bibinfo{author}{Tsai, S.-M.} \& \bibinfo{author}{Stevenson, K.~B.}
\newblock \bibinfo{title}{A new approach to spectroscopic phase curves: {The} emission spectrum of {WASP}-12b observed in quadrature with {HST}/{WFC3}}.
\newblock \emph{\bibinfo{journal}{Astronomy \& Astrophysics}} \textbf{\bibinfo{volume}{646}}, \bibinfo{pages}{A94} (\bibinfo{year}{2021}).

\bibitem{mikal-evans_diurnal_2022}
\bibinfo{author}{Mikal-Evans, T.} \emph{et~al.}
\newblock \bibinfo{title}{Diurnal variations in the stratosphere of the ultrahot giant exoplanet {WASP}-121b}.
\newblock \emph{\bibinfo{journal}{Nature Astronomy}}  (\bibinfo{year}{2022}).

\bibitem{berta_flat_2012}
\bibinfo{author}{Berta, Z.~K.} \emph{et~al.}
\newblock \bibinfo{title}{{THE} {FLAT} {TRANSMISSION} {SPECTRUM} {OF} {THE} {SUPER}-{EARTH} {GJ1214b} {FROM} {WIDE} {FIELD} {CAMERA} 3 {ON} {THE} \textit{{HUBBLE} {SPACE} {TELESCOPE}}}.
\newblock \emph{\bibinfo{journal}{The Astrophysical Journal}} \textbf{\bibinfo{volume}{747}}~(1), \bibinfo{pages}{35} (\bibinfo{year}{2012}).

\bibitem{mandel_analytic_2002}
\bibinfo{author}{Mandel, K.} \& \bibinfo{author}{Agol, E.}
\newblock \bibinfo{title}{Analytic {Light} {Curves} for {Planetary} {Transit} {Searches}}.
\newblock \emph{\bibinfo{journal}{The Astrophysical Journal}} \textbf{\bibinfo{volume}{580}}~(2), \bibinfo{pages}{L171--L175} (\bibinfo{year}{2002}).

\bibitem{carter_empirical_2010}
\bibinfo{author}{Carter, J.~A.} \& \bibinfo{author}{Winn, J.~N.}
\newblock \bibinfo{title}{Empirical {Constraints} on the {Oblateness} of an {Exoplanet}}.
\newblock \emph{\bibinfo{journal}{The Astrophysical Journal}} \textbf{\bibinfo{volume}{709}}~(2), \bibinfo{pages}{1219--1229} (\bibinfo{year}{2010}).
\newblock \bibinfo{note}{ArXiv:0912.1594 [astro-ph]}.

\bibitem{deming_strong_2006}
\bibinfo{author}{Deming, D.}, \bibinfo{author}{Harrington, J.}, \bibinfo{author}{Seager, S.} \& \bibinfo{author}{Richardson, L.~J.}
\newblock \bibinfo{title}{Strong {Infrared} {Emission} from the {Extrasolar} {Planet} {HD} 189733b}.
\newblock \emph{\bibinfo{journal}{The Astrophysical Journal}} \textbf{\bibinfo{volume}{644}}~(1), \bibinfo{pages}{560--564} (\bibinfo{year}{2006}).

\bibitem{kempton_reflective_2023}
\bibinfo{author}{Kempton, E. M.-R.} \emph{et~al.}
\newblock \bibinfo{title}{A reflective, metal-rich atmosphere for {GJ} 1214b from its {JWST} phase curve}.
\newblock \emph{\bibinfo{journal}{Nature}} \textbf{\bibinfo{volume}{620}}~(7972), \bibinfo{pages}{67--71} (\bibinfo{year}{2023}).
\newblock \bibinfo{note}{ArXiv:2305.06240 [astro-ph]}.

\bibitem{bell_eureka_2022}
\bibinfo{author}{Bell, T.~J.} \emph{et~al.}
\newblock \bibinfo{title}{Eureka!: {An} {End}-to-{End} {Pipeline} for {JWST} {Time}-{Series} {Observations}}.
\newblock \emph{\bibinfo{journal}{Journal of Open Source Software}} \textbf{\bibinfo{volume}{7}}~(79), \bibinfo{pages}{4503} (\bibinfo{year}{2022}).
\newblock \bibinfo{note}{ArXiv:2207.03585 [astro-ph]}.

\bibitem{skilling_nested_2006}
\bibinfo{author}{Skilling, J.}
\newblock \bibinfo{title}{Nested sampling for general {Bayesian} computation}.
\newblock \emph{\bibinfo{journal}{Bayesian Analysis}} \textbf{\bibinfo{volume}{1}}~(4), \bibinfo{pages}{833--859} (\bibinfo{year}{2006}).
\newblock \bibinfo{note}{Publisher: International Society for Bayesian Analysis}.

\bibitem{madhusudhan_atmospheric_2018}
\bibinfo{author}{Madhusudhan, N.}
\newblock \bibinfo{title}{ in \textit{Atmospheric {Retrieval} of {Exoplanets}}}  \bibinfo{pages}{2153--2182} (\bibinfo{year}{2018}).
\newblock \urlprefix\url{http://arxiv.org/abs/1808.04824}.
\newblock \bibinfo{note}{ArXiv:1808.04824 [astro-ph]}.

\bibitem{welbanks_application_2023}
\bibinfo{author}{Welbanks, L.}, \bibinfo{author}{McGill, P.}, \bibinfo{author}{Line, M.} \& \bibinfo{author}{Madhusudhan, N.}
\newblock \bibinfo{title}{On the {Application} of {Bayesian} {Leave}-one-out {Cross}-validation to {Exoplanet} {Atmospheric} {Analysis}}.
\newblock \emph{\bibinfo{journal}{The Astronomical Journal}} \textbf{\bibinfo{volume}{165}}~(3), \bibinfo{pages}{112} (\bibinfo{year}{2023}).
\newblock \bibinfo{note}{Publisher: The American Astronomical Society}.

\bibitem{trotta_bayes_2008}
\bibinfo{author}{Trotta, R.}
\newblock \bibinfo{title}{Bayes in the sky: {Bayesian} inference and model selection in cosmology}.
\newblock \emph{\bibinfo{journal}{Contemporary Physics}} \textbf{\bibinfo{volume}{49}}~(2), \bibinfo{pages}{71--104} (\bibinfo{year}{2008}).
\newblock \bibinfo{note}{ArXiv:0803.4089 [astro-ph]}.

\bibitem{benneke_how_2013}
\bibinfo{author}{Benneke, B.} \& \bibinfo{author}{Seager, S.}
\newblock \bibinfo{title}{{HOW} {TO} {DISTINGUISH} {BETWEEN} {CLOUDY} {MINI}-{NEPTUNES} {AND} {WATER}/{VOLATILE}-{DOMINATED} {SUPER}-{EARTHS}}.
\newblock \emph{\bibinfo{journal}{The Astrophysical Journal}} \textbf{\bibinfo{volume}{778}}~(2), \bibinfo{pages}{153} (\bibinfo{year}{2013}).
\newblock \bibinfo{note}{Publisher: The American Astronomical Society}.

\bibitem{welbanks_aurora_2021}
\bibinfo{author}{Welbanks, L.} \& \bibinfo{author}{Madhusudhan, N.}
\newblock \bibinfo{title}{Aurora: {A} {Generalized} {Retrieval} {Framework} for {Exoplanetary} {Transmission} {Spectra}}.
\newblock \emph{\bibinfo{journal}{The Astrophysical Journal}} \textbf{\bibinfo{volume}{913}}~(2), \bibinfo{pages}{114} (\bibinfo{year}{2021}).

\bibitem{fortney_framework_2013}
\bibinfo{author}{Fortney, J.~J.} \emph{et~al.}
\newblock \bibinfo{title}{A {Framework} for {Characterizing} the {Atmospheres} of {Low}-{Mass} {Low}-{Density} {Transiting} {Planets}}.
\newblock \emph{\bibinfo{journal}{arXiv:1306.4329 [astro-ph]}}  (\bibinfo{year}{2013}).
\newblock \bibinfo{note}{ArXiv: 1306.4329}.

\bibitem{line_influence_2016}
\bibinfo{author}{Line, M.~R.} \& \bibinfo{author}{Parmentier, V.}
\newblock \bibinfo{title}{The {Influence} of {Non}-{Uniform} {Cloud} {Cover} on {Transit} {Transmission} {Spectra}}.
\newblock \emph{\bibinfo{journal}{The Astrophysical Journal}} \textbf{\bibinfo{volume}{820}}~(1), \bibinfo{pages}{78} (\bibinfo{year}{2016}).
\newblock \bibinfo{note}{ArXiv: 1511.09443}.

\bibitem{heng_theory_2017}
\bibinfo{author}{Heng, K.} \& \bibinfo{author}{Kitzmann, D.}
\newblock \bibinfo{title}{The theory of transmission spectra revisited: a semi-analytical method for interpreting {WFC3} data and an unresolved challenge}.
\newblock \emph{\bibinfo{journal}{Monthly Notices of the Royal Astronomical Society}} \textbf{\bibinfo{volume}{470}}~(3), \bibinfo{pages}{2972--2981} (\bibinfo{year}{2017}).

\bibitem{welbanks_mass-metallicity_2019}
\bibinfo{author}{Welbanks, L.} \emph{et~al.}
\newblock \bibinfo{title}{Mass-{Metallicity} {Trends} in {Transiting} {Exoplanets} from {Atmospheric} {Abundances} of {H}\$\_2\${O}, {Na}, and {K}}.
\newblock \emph{\bibinfo{journal}{The Astrophysical Journal}} \textbf{\bibinfo{volume}{887}}~(1), \bibinfo{pages}{L20} (\bibinfo{year}{2019}).
\newblock \bibinfo{note}{ArXiv: 1912.04904}.

\bibitem{macdonald_why_2020}
\bibinfo{author}{MacDonald, R.~J.}, \bibinfo{author}{Goyal, J.~M.} \& \bibinfo{author}{Lewis, N.~K.}
\newblock \bibinfo{title}{Why {Is} it {So} {Cold} in {Here}? {Explaining} the {Cold} {Temperatures} {Retrieved} from {Transmission} {Spectra} of {Exoplanet} {Atmospheres}}.
\newblock \emph{\bibinfo{journal}{The Astrophysical Journal}} \textbf{\bibinfo{volume}{893}}~(2), \bibinfo{pages}{L43} (\bibinfo{year}{2020}).

\bibitem{welbanks_atmospheric_2022}
\bibinfo{author}{Welbanks, L.} \& \bibinfo{author}{Madhusudhan, N.}
\newblock \bibinfo{title}{On {Atmospheric} {Retrievals} of {Exoplanets} with {Inhomogeneous} {Terminators}}.
\newblock \emph{\bibinfo{journal}{The Astrophysical Journal}} \textbf{\bibinfo{volume}{933}}~(1), \bibinfo{pages}{79} (\bibinfo{year}{2022}).
\newblock \bibinfo{note}{Publisher: The American Astronomical Society}.

\bibitem{benneke_atmospheric_2012}
\bibinfo{author}{Benneke, B.} \& \bibinfo{author}{Seager, S.}
\newblock \bibinfo{title}{{ATMOSPHERIC} {RETRIEVAL} {FOR} {SUPER}-{EARTHS}: {UNIQUELY} {CONSTRAINING} {THE} {ATMOSPHERIC} {COMPOSITION} {WITH} {TRANSMISSION} {SPECTROSCOPY}}.
\newblock \emph{\bibinfo{journal}{The Astrophysical Journal}} \textbf{\bibinfo{volume}{753}}~(2), \bibinfo{pages}{100} (\bibinfo{year}{2012}).
\newblock \bibinfo{note}{Publisher: The American Astronomical Society}.

\bibitem{guillot_radiative_2010}
\bibinfo{author}{Guillot, T.}
\newblock \bibinfo{title}{On the radiative equilibrium of irradiated planetary atmospheres}.
\newblock \emph{\bibinfo{journal}{Astronomy and Astrophysics}} \textbf{\bibinfo{volume}{520}}, \bibinfo{pages}{A27} (\bibinfo{year}{2010}).

\bibitem{marley_cool_2015}
\bibinfo{author}{Marley, M.~S.} \& \bibinfo{author}{Robinson, T.~D.}
\newblock \bibinfo{title}{On the {Cool} {Side}: {Modeling} the {Atmospheres} of {Brown} {Dwarfs} and {Giant} {Planets}}.
\newblock \emph{\bibinfo{journal}{Annual Review of Astronomy and Astrophysics}} \textbf{\bibinfo{volume}{53}}~(1), \bibinfo{pages}{279--323} (\bibinfo{year}{2015}).
\newblock \bibinfo{note}{ArXiv:1410.6512 [astro-ph]}.

\bibitem{piskorz_ground-_2018}
\bibinfo{author}{Piskorz, D.} \emph{et~al.}
\newblock \bibinfo{title}{Ground- and {Space}-based {Detection} of the {Thermal} {Emission} {Spectrum} of the {Transiting} {Hot} {Jupiter} {KELT}-{2Ab}}.
\newblock \emph{\bibinfo{journal}{The Astronomical Journal}} \textbf{\bibinfo{volume}{156}}~(3), \bibinfo{pages}{133} (\bibinfo{year}{2018}).
\newblock \bibinfo{note}{Publisher: The American Astronomical Society}.

\bibitem{mansfield_unique_2021}
\bibinfo{author}{Mansfield, M.} \emph{et~al.}
\newblock \bibinfo{title}{A unique hot {Jupiter} spectral sequence with evidence for compositional diversity}.
\newblock \emph{\bibinfo{journal}{Nature Astronomy}}  (\bibinfo{year}{2021}).
\newblock \bibinfo{note}{ArXiv: 2110.11272}.

\bibitem{iyer_sphinx_2023}
\bibinfo{author}{Iyer, A.~R.}, \bibinfo{author}{Line, M.~R.}, \bibinfo{author}{Muirhead, P.~S.}, \bibinfo{author}{Fortney, J.~J.} \& \bibinfo{author}{Gharib-Nezhad, E.}
\newblock \bibinfo{title}{The {SPHINX} {M}-dwarf {Spectral} {Grid}. {I}. {Benchmarking} {New} {Model} {Atmospheres} to {Derive} {Fundamental} {M}-dwarf {Properties}}.
\newblock \emph{\bibinfo{journal}{The Astrophysical Journal}} \textbf{\bibinfo{volume}{944}}~(1), \bibinfo{pages}{41} (\bibinfo{year}{2023}).
\newblock \bibinfo{note}{Publisher: The American Astronomical Society}.

\bibitem{tsai_comparative_2021}
\bibinfo{author}{Tsai, S.-M.} \emph{et~al.}
\newblock \bibinfo{title}{A {Comparative} {Study} of {Atmospheric} {Chemistry} with {VULCAN}}.
\newblock \emph{\bibinfo{journal}{The Astrophysical Journal}} \textbf{\bibinfo{volume}{923}}~(2), \bibinfo{pages}{264} (\bibinfo{year}{2021}).

\bibitem{tsai_vulcan_2017}
\bibinfo{author}{Tsai, S.-M.} \emph{et~al.}
\newblock \bibinfo{title}{{VULCAN} : {An} {Open}-source, {Validated} {Chemical} {Kinetics} {Python} {Code} for {Exoplanetary} {Atmospheres}}.
\newblock \emph{\bibinfo{journal}{The Astrophysical Journal Supplement Series}} \textbf{\bibinfo{volume}{228}}~(2), \bibinfo{pages}{20} (\bibinfo{year}{2017}).

\bibitem{tsai_photochemically_2023}
\bibinfo{author}{Tsai, S.-M.} \emph{et~al.}
\newblock \bibinfo{title}{Photochemically produced {SO2} in the atmosphere of {WASP}-39b}.
\newblock \emph{\bibinfo{journal}{Nature}} \textbf{\bibinfo{volume}{617}}~(7961), \bibinfo{pages}{483--487} (\bibinfo{year}{2023}).
\newblock \bibinfo{note}{Number: 7961 Publisher: Nature Publishing Group}.

\bibitem{carter_jwst_2023}
\bibinfo{author}{Carter, A.~L.} \emph{et~al.}
\newblock \bibinfo{title}{The {JWST} {Early} {Release} {Science} {Program} for {Direct} {Observations} of {Exoplanetary} {Systems} {I}: {High}-contrast {Imaging} of the {Exoplanet} {HIP} 65426 b from 2 to 16 $\mu$m}.
\newblock \emph{\bibinfo{journal}{The Astrophysical Journal Letters}} \textbf{\bibinfo{volume}{951}}~(1), \bibinfo{pages}{L20} (\bibinfo{year}{2023}).
\newblock \bibinfo{note}{Publisher: The American Astronomical Society}.

\bibitem{husser_new_2013}
\bibinfo{author}{Husser, T.-O.} \emph{et~al.}
\newblock \bibinfo{title}{A new extensive library of {PHOENIX} stellar atmospheres and synthetic spectra}.
\newblock \emph{\bibinfo{journal}{Astronomy \& Astrophysics}} \textbf{\bibinfo{volume}{553}}, \bibinfo{pages}{A6} (\bibinfo{year}{2013}).

\bibitem{lodders_44_2009}
\bibinfo{author}{Lodders, K.}, \bibinfo{author}{Palme, H.} \& \bibinfo{author}{Gail, H.-P.}
\newblock \bibinfo{title}{ in \textit{4.4 {Abundances} of the elements in the {Solar} {System}}} (ed.\bibinfo{editor}{Trümper, J.}) \emph{\bibinfo{booktitle}{Solar {System}}}, Vol.~\bibinfo{volume}{4B} \bibinfo{pages}{712--770} (\bibinfo{publisher}{Springer Berlin Heidelberg}, \bibinfo{address}{Berlin, Heidelberg}, \bibinfo{year}{2009}).
\newblock \urlprefix\url{http://materials.springer.com/lb/docs/sm_lbs_978-3-540-88055-4_34}.
\newblock \bibinfo{note}{Series Title: Landolt-Börnstein - Group VI Astronomy and Astrophysics}.

\bibitem{gordon_computer_1994}
\bibinfo{author}{Gordon, S.} \& \bibinfo{author}{Mcbride, B.~J.}
\newblock \bibinfo{title}{Computer program for calculation of complex chemical equilibrium compositions and applications. {Part} 1: {Analysis}} (\bibinfo{year}{1994}).
\newblock \urlprefix\url{https://ntrs.nasa.gov/citations/19950013764}.
\newblock \bibinfo{note}{NTRS Author Affiliations: NASA Lewis Research Center NTRS Report/Patent Number: NAS 1.61:1311 NTRS Document ID: 19950013764 NTRS Research Center: Legacy CDMS (CDMS)}.

\bibitem{moses_disequilibrium_2011}
\bibinfo{author}{Moses, J.~I.} \emph{et~al.}
\newblock \bibinfo{title}{{DISEQUILIBRIUM} {CARBON}, {OXYGEN}, {AND} {NITROGEN} {CHEMISTRY} {IN} {THE} {ATMOSPHERES} {OF} {HD} 189733b {AND} {HD} 209458b}.
\newblock \emph{\bibinfo{journal}{The Astrophysical Journal}} \textbf{\bibinfo{volume}{737}}~(1), \bibinfo{pages}{15} (\bibinfo{year}{2011}).

\bibitem{thorngren_intrinsic_2019}
\bibinfo{author}{Thorngren, D.}, \bibinfo{author}{Gao, P.} \& \bibinfo{author}{Fortney, J.~J.}
\newblock \bibinfo{title}{The {Intrinsic} {Temperature} and {Radiative}–{Convective} {Boundary} {Depth} in the {Atmospheres} of {Hot} {Jupiters}}.
\newblock \emph{\bibinfo{journal}{The Astrophysical Journal Letters}} \textbf{\bibinfo{volume}{884}}~(1), \bibinfo{pages}{L6} (\bibinfo{year}{2019}).
\newblock \bibinfo{note}{Publisher: The American Astronomical Society}.

\bibitem{karman_update_2019}
\bibinfo{author}{Karman, T.} \emph{et~al.}
\newblock \bibinfo{title}{Update of the {HITRAN} collision-induced absorption section}.
\newblock \emph{\bibinfo{journal}{Icarus}} \textbf{\bibinfo{volume}{328}}, \bibinfo{pages}{160--175} (\bibinfo{year}{2019}).

\bibitem{polyansky_exomol_2018}
\bibinfo{author}{Polyansky, O.~L.} \emph{et~al.}
\newblock \bibinfo{title}{{ExoMol} molecular line lists {XXX}: a complete high-accuracy line list for water}.
\newblock \emph{\bibinfo{journal}{Monthly Notices of the Royal Astronomical Society}} \textbf{\bibinfo{volume}{480}}~(2), \bibinfo{pages}{2597--2608} (\bibinfo{year}{2018}).

\bibitem{huang__reliable_2014}
\bibinfo{author}{Huang, X.}, \bibinfo{author}{Gamache, R.~R.}, \bibinfo{author}{Freedman, R.~S.}, \bibinfo{author}{Schwenke, D.~W.} \& \bibinfo{author}{Lee, T.~J.}
\newblock \bibinfo{title}{Reliable infrared line lists for 13 {CO2} isotopologues up to E=18,000cm$^{-1}$ and {1500K}, with line shape parameters}.
\newblock \emph{\bibinfo{journal}{Journal of Quantitative Spectroscopy and Radiative Transfer}} \textbf{\bibinfo{volume}{147}}, \bibinfo{pages}{134--144} (\bibinfo{year}{2014}).

\bibitem{allard_kh_2016}
\bibinfo{author}{Allard, N.~F.}, \bibinfo{author}{Spiegelman, F.} \& \bibinfo{author}{Kielkopf, J.~F.}
\newblock \bibinfo{title}{K–{H} $_{\textrm{2}}$ line shapes for the spectra of cool brown dwarfs}.
\newblock \emph{\bibinfo{journal}{Astronomy \& Astrophysics}} \textbf{\bibinfo{volume}{589}}, \bibinfo{pages}{A21} (\bibinfo{year}{2016}).

\bibitem{underwood_exomol_2016}
\bibinfo{author}{Underwood, D.~S.} \emph{et~al.}
\newblock \bibinfo{title}{{ExoMol} molecular line lists – {XIV}. {The} rotation–vibration spectrum of hot {SO2}}.
\newblock \emph{\bibinfo{journal}{Monthly Notices of the Royal Astronomical Society}} \textbf{\bibinfo{volume}{459}}~(4), \bibinfo{pages}{3890--3899} (\bibinfo{year}{2016}).

\bibitem{hargreaves_accurate_2020}
\bibinfo{author}{Hargreaves, R.~J.} \emph{et~al.}
\newblock \bibinfo{title}{An {Accurate}, {Extensive}, and {Practical} {Line} {List} of {Methane} for the {HITEMP} {Database}}.
\newblock \emph{\bibinfo{journal}{The Astrophysical Journal Supplement Series}} \textbf{\bibinfo{volume}{247}}~(2), \bibinfo{pages}{55} (\bibinfo{year}{2020}).
\newblock \bibinfo{note}{Publisher: The American Astronomical Society}.

\bibitem{coles_exomol_2019}
\bibinfo{author}{Coles, P.~A.}, \bibinfo{author}{Yurchenko, S.~N.} \& \bibinfo{author}{Tennyson, J.}
\newblock \bibinfo{title}{{ExoMol} molecular line lists – {XXXV}. {A} rotation-vibration line list for hot ammonia}.
\newblock \emph{\bibinfo{journal}{Monthly Notices of the Royal Astronomical Society}} \textbf{\bibinfo{volume}{490}}~(4), \bibinfo{pages}{4638--4647} (\bibinfo{year}{2019}).

\bibitem{harris_improved_2006}
\bibinfo{author}{Harris, G.~J.}, \bibinfo{author}{Tennyson, J.}, \bibinfo{author}{Kaminsky, B.~M.}, \bibinfo{author}{Pavlenko, Y.~V.} \& \bibinfo{author}{Jones, H. R.~A.}
\newblock \bibinfo{title}{Improved {HCN}/{HNC} linelist, model atmospheres and synthetic spectra for {WZ} {Cas}}.
\newblock \emph{\bibinfo{journal}{Monthly Notices of the Royal Astronomical Society}} \textbf{\bibinfo{volume}{367}}~(1), \bibinfo{pages}{400--406} (\bibinfo{year}{2006}).

\bibitem{chubb_exomol_2020}
\bibinfo{author}{Chubb, K.~L.}, \bibinfo{author}{Tennyson, J.} \& \bibinfo{author}{Yurchenko, S.~N.}
\newblock \bibinfo{title}{{ExoMol} molecular line lists – {XXXVII}. {Spectra} of acetylene}.
\newblock \emph{\bibinfo{journal}{Monthly Notices of the Royal Astronomical Society}} \textbf{\bibinfo{volume}{493}}~(2), \bibinfo{pages}{1531--1545} (\bibinfo{year}{2020}).

\bibitem{azzam_exomol_2016}
\bibinfo{author}{Azzam, A. A.~A.}, \bibinfo{author}{Tennyson, J.}, \bibinfo{author}{Yurchenko, S.~N.} \& \bibinfo{author}{Naumenko, O.~V.}
\newblock \bibinfo{title}{{ExoMol} molecular line lists – {XVI}. {The} rotation–vibration spectrum of hot {H2S}}.
\newblock \emph{\bibinfo{journal}{Monthly Notices of the Royal Astronomical Society}} \textbf{\bibinfo{volume}{460}}~(4), \bibinfo{pages}{4063--4074} (\bibinfo{year}{2016}).

\bibitem{virtanen_scipy_2020}
\bibinfo{author}{Virtanen, P.} \emph{et~al.}
\newblock \bibinfo{title}{{SciPy} 1.0: fundamental algorithms for scientific computing in {Python}}.
\newblock \emph{\bibinfo{journal}{Nature Methods}} \textbf{\bibinfo{volume}{17}}~(3), \bibinfo{pages}{261--272} (\bibinfo{year}{2020}).
\newblock \bibinfo{note}{Number: 3 Publisher: Nature Publishing Group}.

\bibitem{etangs_rayleigh_2008}
\bibinfo{author}{Etangs, A. L.~d.}, \bibinfo{author}{Pont, F.}, \bibinfo{author}{Vidal-Madjar, A.} \& \bibinfo{author}{Sing, D.}
\newblock \bibinfo{title}{Rayleigh scattering in the transit spectrum of {HD} 189733b}.
\newblock \emph{\bibinfo{journal}{Astronomy \& Astrophysics}} \textbf{\bibinfo{volume}{481}}~(2), \bibinfo{pages}{L83--L86} (\bibinfo{year}{2008}).
\newblock \bibinfo{note}{Number: 2 Publisher: EDP Sciences}.

\bibitem{fisher_how_2022}
\bibinfo{author}{Fisher, C.} \& \bibinfo{author}{Heng, K.}
\newblock \bibinfo{title}{How {Do} {We} {Optimally} {Sample} {Model} {Grids} of {Exoplanet} {Spectra}?}
\newblock \emph{\bibinfo{journal}{The Astrophysical Journal}} \textbf{\bibinfo{volume}{934}}~(1), \bibinfo{pages}{31} (\bibinfo{year}{2022}).
\newblock \bibinfo{note}{Publisher: The American Astronomical Society}.

\bibitem{line_systematic_2013}
\bibinfo{author}{Line, M.~R.} \emph{et~al.}
\newblock \bibinfo{title}{A {Systematic} {Retrieval} {Analysis} of {Secondary} {Eclipse} {Spectra} {I}: {A} {Comparison} of {Atmospheric} {Retrieval} {Techniques}}.
\newblock \emph{\bibinfo{journal}{The Astrophysical Journal}} \textbf{\bibinfo{volume}{775}}~(2), \bibinfo{pages}{137} (\bibinfo{year}{2013}).
\newblock \bibinfo{note}{ArXiv: 1304.5561}.

\bibitem{kreidberg_detection_2015}
\bibinfo{author}{Kreidberg, L.} \emph{et~al.}
\newblock \bibinfo{title}{A detection of water in the transmission spectrum of the hot {Jupiter} {WASP}-12b and implications for its atmospheric composition}.
\newblock \emph{\bibinfo{journal}{The Astrophysical Journal}} \textbf{\bibinfo{volume}{814}}~(1), \bibinfo{pages}{66} (\bibinfo{year}{2015}).

\bibitem{kreidberg_water_2018}
\bibinfo{author}{Kreidberg, L.}, \bibinfo{author}{Line, M.~R.}, \bibinfo{author}{Thorngren, D.}, \bibinfo{author}{Morley, C.~V.} \& \bibinfo{author}{Stevenson, K.~B.}
\newblock \bibinfo{title}{Water, {High}-altitude {Condensates}, and {Possible} {Methane} {Depletion} in the {Atmosphere} of the {Warm} {Super}-{Neptune} {WASP}-107b}.
\newblock \emph{\bibinfo{journal}{The Astrophysical Journal}} \textbf{\bibinfo{volume}{858}}~(1), \bibinfo{pages}{L6} (\bibinfo{year}{2018}).

\bibitem{parmentier_non-grey_2014}
\bibinfo{author}{Parmentier, V.} \& \bibinfo{author}{Guillot, T.}
\newblock \bibinfo{title}{A non-grey analytical model for irradiated atmospheres: {I}. {Derivation}}.
\newblock \emph{\bibinfo{journal}{Astronomy \& Astrophysics}} \textbf{\bibinfo{volume}{562}}, \bibinfo{pages}{A133} (\bibinfo{year}{2014}).

\bibitem{sing_observational_2018}
\bibinfo{author}{Sing, D.~K.}
\newblock \bibinfo{title}{Observational {Techniques} {With} {Transiting} {Exoplanetary} {Atmospheres}} (\bibinfo{year}{2018}).
\newblock \urlprefix\url{http://arxiv.org/abs/1804.07357}.
\newblock \bibinfo{note}{ArXiv:1804.07357 [astro-ph]}.

\bibitem{henry_techniques_1999}
\bibinfo{author}{Henry, G.~W.}
\newblock \bibinfo{title}{Techniques for {Automated} {High}‐{Precision} {Photometry} of {Sun}‐like {Stars}}.
\newblock \emph{\bibinfo{journal}{Publications of the Astronomical Society of the Pacific}} \textbf{\bibinfo{volume}{111}}~(761), \bibinfo{pages}{845} (\bibinfo{year}{1999}).
\newblock \bibinfo{note}{Publisher: The University of Chicago Press}.

\bibitem{vanicek_further_1971}
\bibinfo{author}{Vaníček, P.}
\newblock \bibinfo{title}{Further development and properties of the spectral analysis by least-squares}.
\newblock \emph{\bibinfo{journal}{Astrophysics and Space Science}} \textbf{\bibinfo{volume}{12}}~(1), \bibinfo{pages}{10--33} (\bibinfo{year}{1971}).

\bibitem{henry_nine_2022}
\bibinfo{author}{Henry, G.~W.}, \bibinfo{author}{Fekel, F.~C.} \& \bibinfo{author}{Williamson, M.~H.}
\newblock \bibinfo{title}{Nine {Bright} $\gamma$ {Doradus} {Variables} {Discovered} with {Ground}-based {Photometry}}.
\newblock \emph{\bibinfo{journal}{The Astronomical Journal}} \textbf{\bibinfo{volume}{163}}~(4), \bibinfo{pages}{180} (\bibinfo{year}{2022}).
\newblock \bibinfo{note}{Publisher: The American Astronomical Society}.

\bibitem{harris_array_2020}
\bibinfo{author}{Harris, C.~R.} \emph{et~al.}
\newblock \bibinfo{title}{Array programming with {NumPy}}.
\newblock \emph{\bibinfo{journal}{Nature}} \textbf{\bibinfo{volume}{585}}~(7825), \bibinfo{pages}{357--362} (\bibinfo{year}{2020}).
\newblock \bibinfo{note}{Publisher: Nature Publishing Group}.

\bibitem{hunter_matplotlib_2007}
\bibinfo{author}{Hunter, J.~D.}
\newblock \bibinfo{title}{Matplotlib: {A} {2D} {Graphics} {Environment}}.
\newblock \emph{\bibinfo{journal}{Computing in Science \& Engineering}} \textbf{\bibinfo{volume}{9}}~(3), \bibinfo{pages}{90--95} (\bibinfo{year}{2007}).
\newblock \bibinfo{note}{Conference Name: Computing in Science \& Engineering}.

\end{thebibliography}


\newpage
\backmatter
\section*{Methods} \label{sec:methods}
\setcounter{page}{1}
\setcounter{figure}{0}
\renewcommand{\figurename}{Extended Data Fig.}
\renewcommand{\tablename}{Extended Data Table}
\subsection*{Data analysis}

We performed four independent data analysis routines on the F444W and F322W2 LW datasets from spectral extraction to light curve fitting. Each analysis will be described in detail in the following. 

\subsection*{Analysis A}

G. Fu started the data reduction with obtaining the \texttt{uncal.fits} images from the \texttt{MAST} and then processed by the default JWST pipeline with \texttt{steps.jump} turned off in stage one to produce the \texttt{rateints.fits} images. Next, we performed background row-by-row background subtraction \citeApp{schlawin_jwst_2020} for each integration by taking the row median of the unilluminated region (columns 0 to 512 for F444 and 1850 to 2048 for F322W2) and subtracting it from each row. We then cross-correlated the spectral line spread function with every column to get the vertical position of the spectral trace for each column. A third-order polynomial is then fitted to the vertical position versus column number to determine the spectral trace location for every integration. The spectrum is extracted with a 6-pixel wide aperture centered on the fitted polynomial. Various extraction aperture sizes were tested and a 6-pixel aperture resulted in the lowest scattering in the light curve. The short exposure time per integration and a high number of integrations per visit means any cosmic ray or hot pixel has negligible effects on the transit light curve, in addition, we flagged and removed any outliers ($>$5$\sigma$) based on the scatter of the transit light curve baseline for every column. All wavelength channels are then summed to create the whitelight curve. 

The whitelight curve fitting uses a combination of \texttt{Batman} \citeApp{kreidberg_batman_2015} and \texttt{emcee} \citeApp{foreman-mackey_emcee_2013} with nine parameters including transit-mid time, planet radius, linear slope in time, constant scaling factor, two exponential ramp coefficients, a/R$_\star$, inclination and the second quadratic limb darkening coefficient. We used the 3D stellar models \citeApp{magic_stagger-grid_2015} for determining the quadratic limb darkening parameters \citeApp{grant_exo-ticexotic-ld_2022} and then fixed the first quadratic term \citeApp{ahrer_early_2022}. Our best-fit system parameters including inclination (85.714$\pm$0.01 deg) and a/R$_\star$ (8.88$\pm$0.01) agree with \citeApp{agol_climate_2010} to within 1 sigma uncertainty. We excluded the starspot crossing region (Figure \ref{fig:wl}) in the whitelight fit for the F322W2 visit. The starspot residuals are then fitted with a simple second-order polynomial to model the feature's shape.

We fit the spectroscopic light curves with \texttt{Batman} and \texttt{scipy.optimize.curvefit}. There are six free-fitting parameters including planet radius, linear slope in time, constant scaling factor, two exponential ramp coefficients, and the second quadratic limb darkening coefficient. The one standard deviation uncertainties are calculated by taking the square root of the diagonals of the covariance matrix for all fitting parameters.  For the F322W2 visit, there is another additional scaling parameter for the starspot crossing region. We scale the polynomial starspot feature model from the whitelight residual to fit the starspot feature in each wavelength channel \citeApp{sing_hubble_2011, fu_water_2022}. The mid-transit times, a/R$_\star$, and inclination are fixed to the corresponding best-fit values from the whitelight. We used the wavelength solutions from the Flight Calibration Program 1076.

\begin{figure*}
\centering
  \includegraphics[width=\textwidth,keepaspectratio]{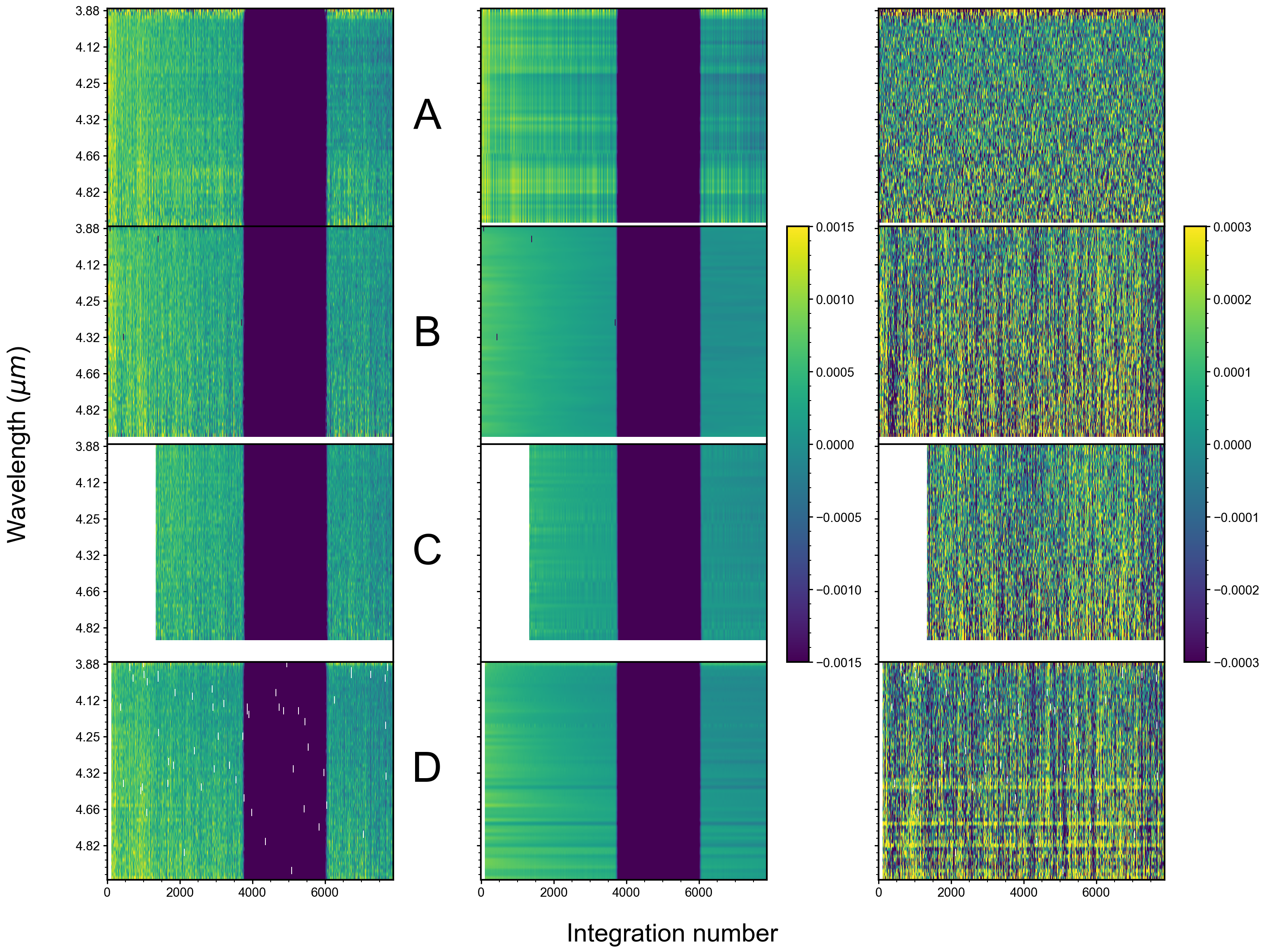}
  \caption{\textbf{F444W data analysis products.} Extracted light curves (left), best-fit models (mid), and residuals (right) for every F444W wavelength channel from four different independent reduction pipelines.}
  \label{fig:F444_data_model_residual_each_amp}
\end{figure*}

\begin{figure*}
\centering
  \includegraphics[width=\textwidth,keepaspectratio]{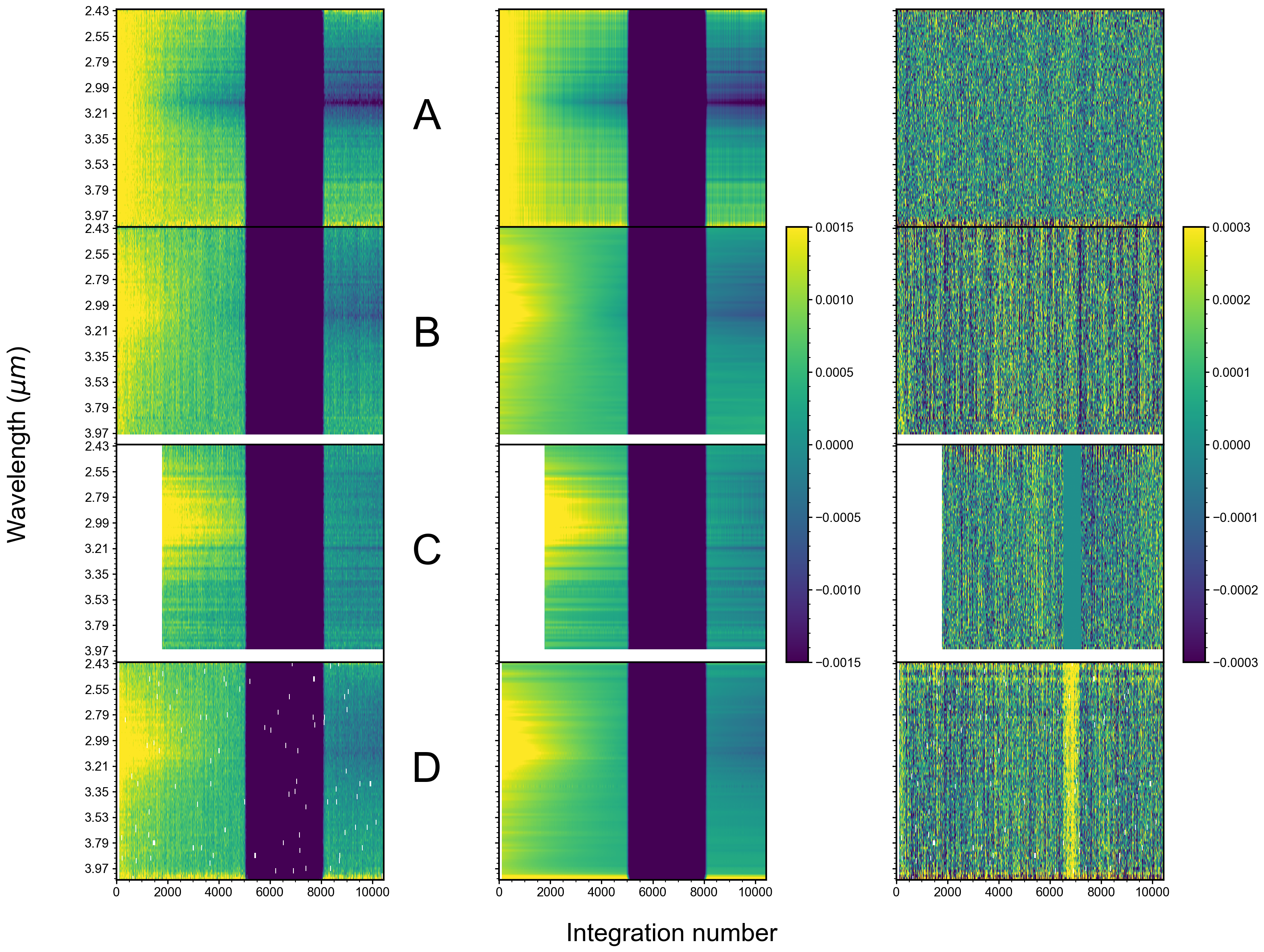}
  \caption{\textbf{F322W2 data analysis products.} Extracted light curves (left), best-fit models (mid), and residuals (right) for every F322W2 wavelength channel from four different independent reduction pipelines.}
  \label{fig:F322_data_model_residual_each_amp}
\end{figure*}

\begin{figure*}
\centering
  \includegraphics[width=\textwidth,keepaspectratio]{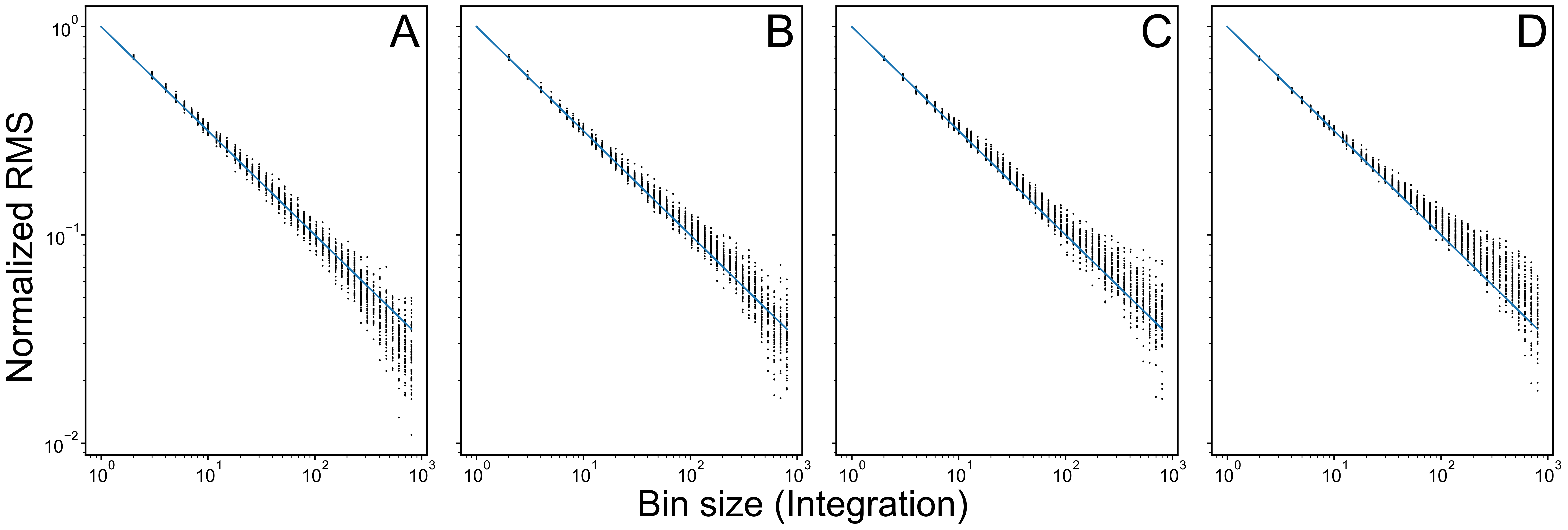}
  \caption{\textbf{Allan variance plot of the F444W residuals from all four reductions.} Extracted light curves (left), best-fit models (mid), and residuals (right) for every F444W wavelength channel from four different independent reduction pipelines.}
  \label{fig:allen_plot_F444_each_amp}
\end{figure*}

\begin{figure*}
\centering
  \includegraphics[width=\textwidth,keepaspectratio]{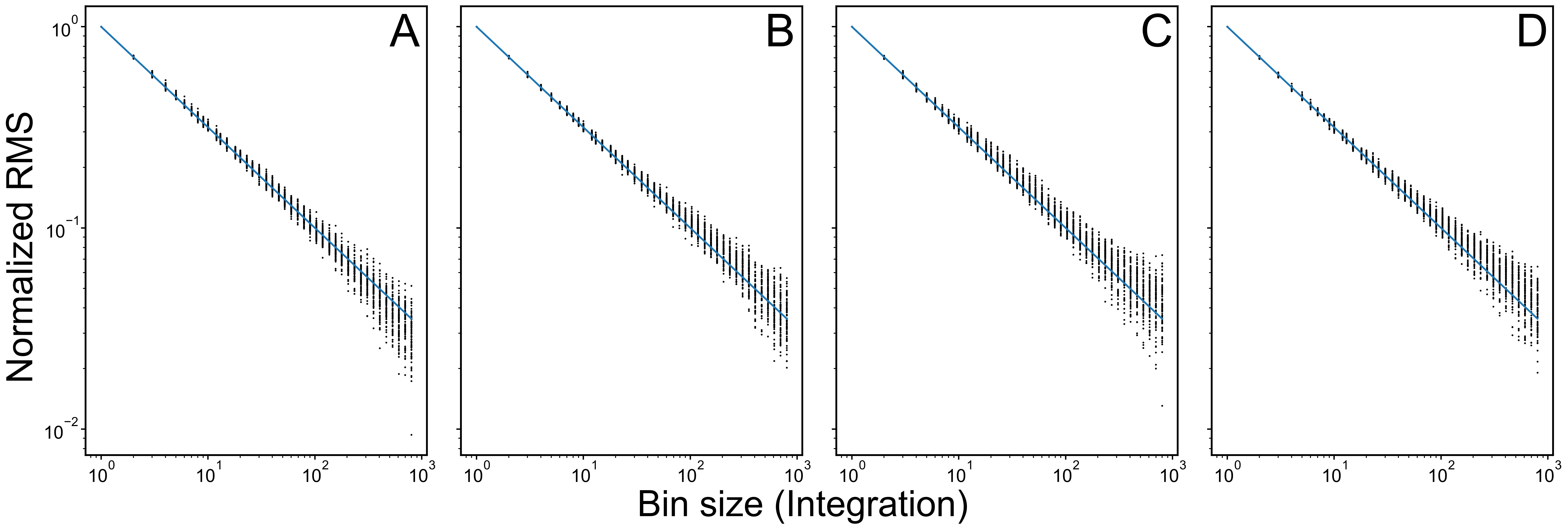}
  \caption{\textbf{Allan variance plot of the F322W2 residuals from all four reductions.} Extracted light curves (left), best-fit models (mid), and residuals (right) for every F322W2 wavelength channel from four different independent reduction pipelines.}
  \label{fig:allen_plot_F322_each_amp}
\end{figure*}

\begin{figure*}
\centering
  \includegraphics[width=\textwidth,keepaspectratio]{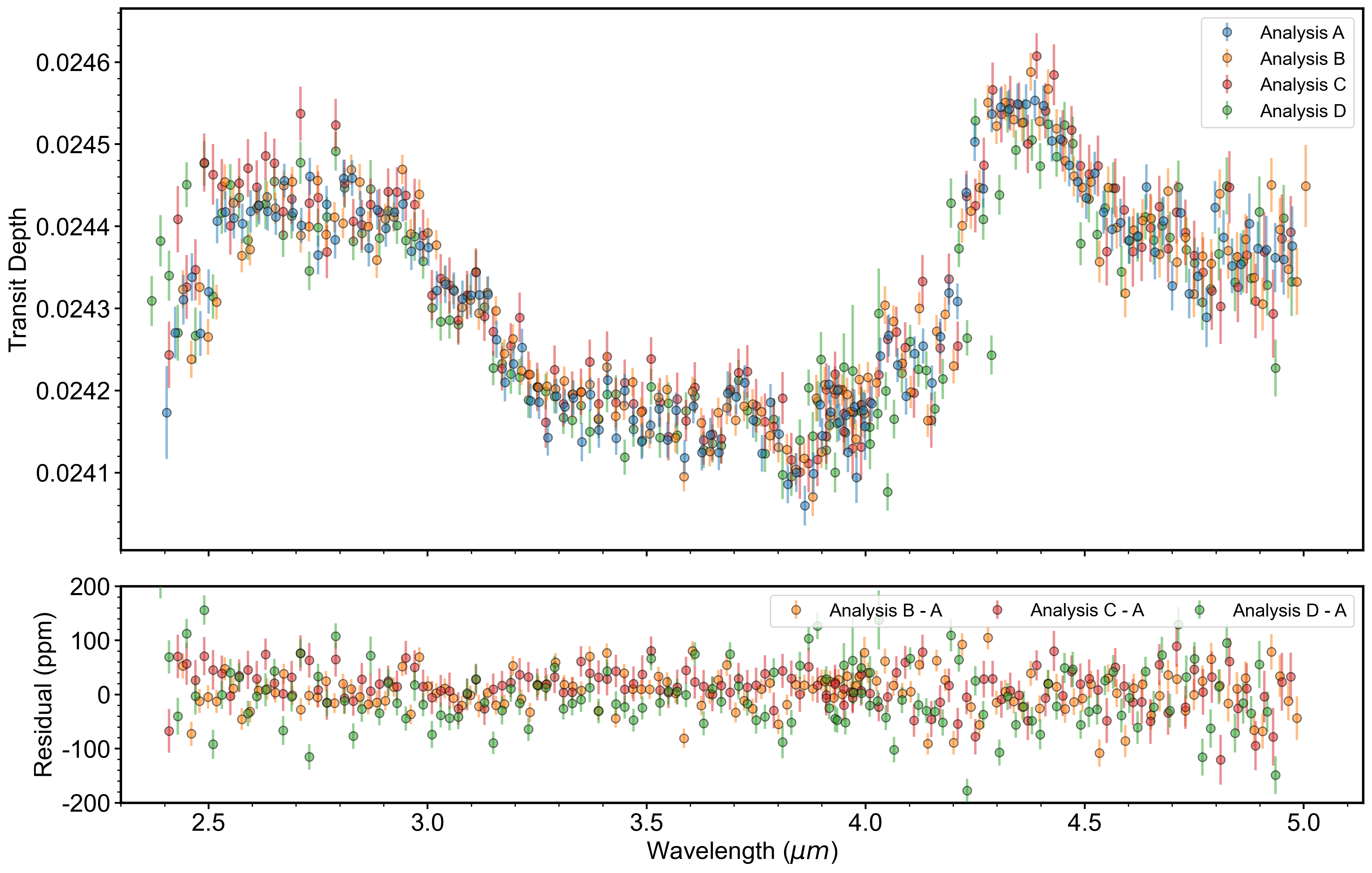}
  \caption{\textbf{NIRCam transmission spectra of HD 189733b from the four independent reductions.} Comparison of the spectra from the four data analyses before applying stellar heterogeneity and nightside emission contamination corrections.}
  \label{fig:compare_reductions}
\end{figure*}

\begin{figure*}
\centering
  \includegraphics[width=\textwidth,keepaspectratio]{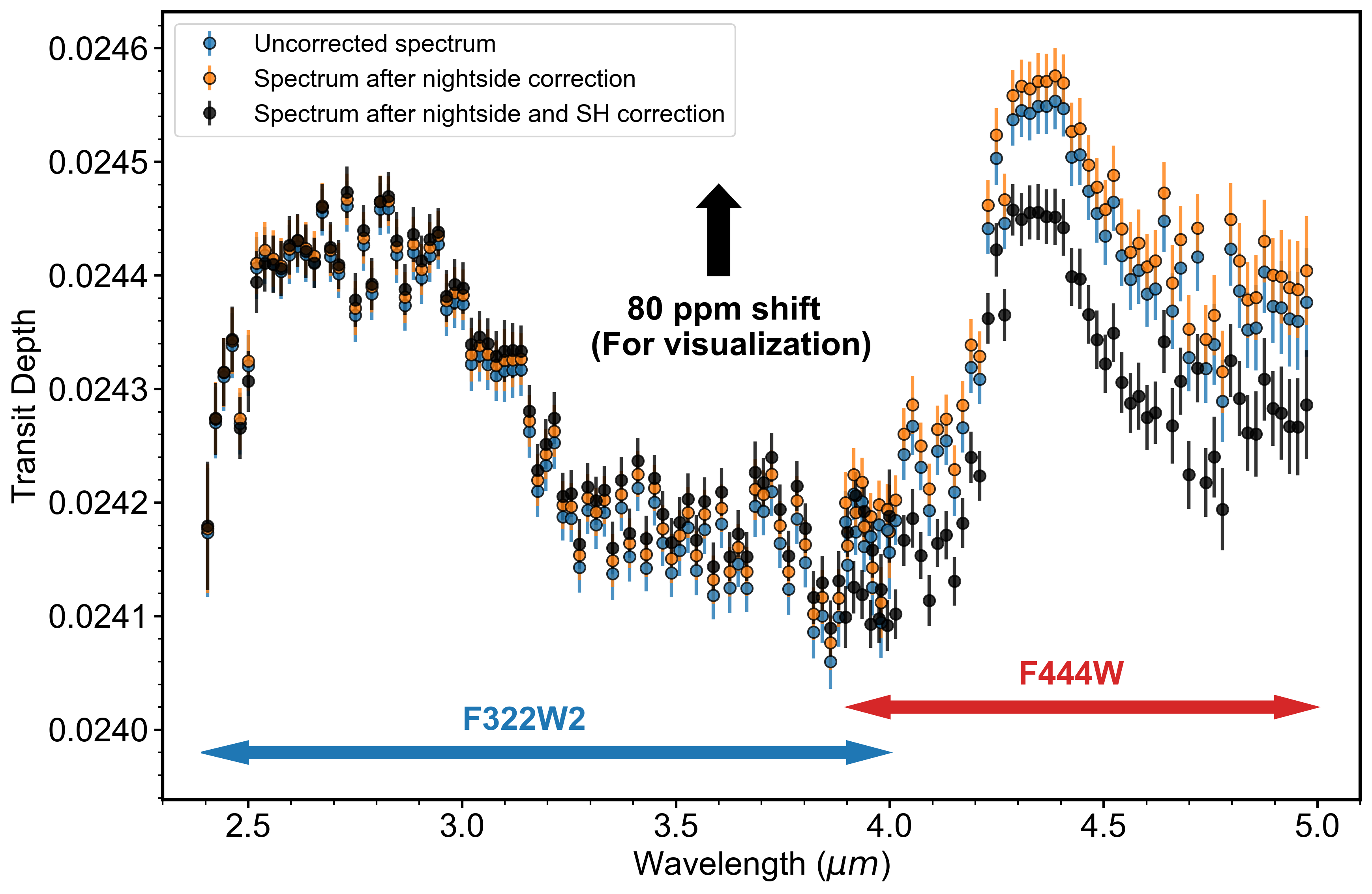}
  \caption{\textbf{The effect of nightside dilution and stellar heterogeneity correction on the spectrum.} The uncorrected spectrum from analysis A is shown in blue. The nightside dilution corrected spectrum is in orange. The nightside dilution and stellar heterogeneity correction applied spectrum is in black. It has been shifted up with a constant offset of 80 ppm to better demonstrate the wavelength-dependent changes. An offset parameter between F322W2 and F444W spectra is joint fitted in the atmospheric modeling and the best-fit offset value is $\sim$20ppm which is consistent to within 1$\sigma$ between spectra from the overlapped wavelength region.}
  \label{fig:correction_plot}
\end{figure*}

Each NIRCam detector has 4 amplifiers and the spectrum is dispersed horizontally across the different amplifiers \citeApp{schlawin_jwst_2020}. The fast readout direction is also along the wavelength dispersion direction. This instrument setup combination introduces amplifier-specific correlated 1/f readout noise in the wavelength direction. Ideally, this noise can be subtracted if there are unilluminated pixels for every row in each amplifier. However, since the spectra are dispersed continuously across the different amplifiers, this correction cannot be done for the entire spectrum. Therefore, we see leftover 1/f noise residuals that are correlated across wavelengths in light curve fits (Figure \ref{fig:F444_data_model_residual_each_amp} and \ref{fig:F322_data_model_residual_each_amp}). We noticed these high-frequency residual features (vertical stripes right column of Figure \ref{fig:F444_data_model_residual_each_amp} and \ref{fig:F322_data_model_residual_each_amp}) are mostly shared across all amplifiers with some minor features shared within each amplifier. To better remove the 1/f noise, we applied an amplifier-specific common-mode correction \citeApp{arcangeli_new_2021, mikal-evans_diurnal_2022, berta_flat_2012} which leverages the fact that light curve systematic structures are shared across different wavelength, especially within each amplifier. We first fit the whitelight for each amplifier and then use its residual as a free-fitting parameter scaled linearly for each spectroscopic channel light curve fit within that amplifier. The improvement can be seen in figure \ref{fig:F444_data_model_residual_each_amp}, \ref{fig:F322_data_model_residual_each_amp}, \ref{fig:allen_plot_F444_each_amp} and \ref{fig:allen_plot_F322_each_amp} where the residuals of method A do not show the strong vertical stripes compared to other reductions where common-mode correction is not used.

\subsection*{Analysis B}

D. Deming begins his version of the data analysis using the \texttt{rateints.fits} files processed by the default JWST pipeline with \texttt{steps.jump} turned off in stage one. For both the F322W2 and F444W bands, we use the same process, albeit with different stellar limb darkening coefficients (LDCs).  For each rateints file, we converted the Julian date (BJD/TDB) of each integration to a tentative orbital phase using the timing ephemeris from Baluev et al. (2019).  We adopt preliminary quadratic LDCs for the star from the Exoplanet Characterization Toolkit (https://exoctk.stsci.edu), using two wavelengths that are median values over the F322 and F444 bands.  The quadratic coefficient is subsequently adjusted by the fitting process at each individual wavelength (single detector column), as described below.

Extracting a spectrum for each integration proceeds by finding the spatial peak of the spectrum in each column. We find that peak by fitting a Gaussian function to the stellar intensity spatial profile at each wavelength, after subtracting the background. We smooth the spatial centroid using a 9-column median in order to follow the curvature of the spectrum while avoiding outlying values.  We experimented with different algorithms for background subtraction. Given that this star is very bright, background subtraction has a very minor effect, and we settled on subtracting a simple per-row offset that was determined by a median over columns where the stellar flux is negligible (columns 0 to 512 for F444 and 1850 to 2048 for F322W2).  With the spatial center of the spectrum determined at each wavelength, we sum the central pixel plus $\pm$5-pixels above and below the central pixel to define the stellar flux at that wavelength and orbital phase.  For consistency between different versions of the data analysis, we adopted the same wavelength calibration (wavelengths versus detector columns) as in the analysis by G. Fu.

With the stellar flux determined at each wavelength and orbital phase, we construct a sum of all wavelengths and make a correction to the transit ephemeris so that the summed "white-light" transit curve is accurately centered on a phase equal to zero.  We then perform a preliminary wavelength-by-wavelength fit of monochromatic transit curves, using the transit model from Ref. \citeApp{mandel_analytic_2002}, adopting the geometric orbital parameters from Ref. \citeApp{carter_empirical_2010}, and using the preliminary LDCs described above.  We fit the transit curve to the stellar flux value using a multi-variate linear regression and varying the depth of the transit as well as spot-crossing and baseline parameters (see below). Calculating the difference between the data and the fitted transit curve at each wavelength, we zero-weight any outlying data points that differ from the fitted curve by more than $3.5\sigma$, where we estimate $\sigma$ from the observed point-to-point scatter at each wavelength.

After zero-weighting outlying data points at each wavelength, we re-do the transit fit.  This final fitting process varies the transit depth (i.e., planet radius), the quadratic (but not linear) LDC, and also a possible wavelength-to-wavelength variation in the central phase and duration of the transit (concluded to be negligible). Both the initial and final fits at each wavelength also include a perturbation to the transit curve due to the star-spot crossing, and we adopt the same phase range for that spot crossing as used by G. Fu (-0.0005947 to +0.0068871). The spot crossing is detected only in the F322 band, but we allow for it in the F444 fitting process for consistency and to verify that it is absent at that wavelength. Both the initial and final fits of the transit curve also include a baseline "ramp" \citeApp{deming_strong_2006} that uses a quadratic term in the orbital phase.  The final fit also adds an exponential term in the ramp function that is intended to account for a transient effect in the baseline at the start of the observations. This final fit at each wavelength is accomplished by a gradient-search algorithm ("curvefit" in IDL) that finds the minimum chi-squared.

We estimated errors on the planet radius using a bootstrap procedure.  Having determined the best-fit transit curve at each wavelength, and the standard deviation of the point-to-point temporal scatter, we make 100 synthetic transit curves using Gaussian noise added to the best-fit curve.  Re-fitting those 100 synthetic transits, we calculate the standard deviation of the resulting planet radius, and we adopt that value as the error bar for the planet radius at each wavelength.  Also, we verified that the temporal scatter at each wavelength is closely approximately by a Gaussian function, hence our adoption of Gaussian noise for this bootstrap process is appropriate.

\subsection*{Analysis C}

M. Zhang's NIRCam data reduction was done using SPARTA, a completely independent pipeline that does not use any code from any other pipeline. The MIRI portion of SPARTA was first described in \citeApp{kempton_reflective_2023}. The NIRCam data is analyzed in a very similar way to the MIRI data. In the first stage, we subtract the superbias, subtract the reference pixels, apply non-linearity correction, subtract the dark, and fit for the up-the-ramp reads.  In the second stage, we remove the background. We first attempt to remove 1/f noise to the maximum extent possible by subtracting the median of unilluminated columns (4-600 for F444W, 1894-2044 for F322W2) of each row, from the entire row. We then remove any wavelength-dependent background by subtracting the median of the background region of each column, from the entire column. The background region is defined to be rows 4-11 (the first 4 are reference pixels) and 57-64, i.e. the 7 pixels closest to the top and bottom edges. In the third stage, we perform sum extraction. Unlike in \citeApp{kempton_reflective_2023}, where we sum the pixels within a predefined rectangular window, in this reduction we identify the trace and define the window based on it. For each column, we fit a Gaussian to the spatial profile to identify the location of the trace; we then fit a second-degree polynomial to the trace location as a function of column number. We use this polynomial to define extraction windows with a half-width of 6 pixels, and take into account partial pixels when summing the flux within the window. In the fifth stage, we gather all the spectra, compute normalized spectroscopic light curves, and mask $>$4-sigma outliers in each spectroscopic light curve.

After obtaining the spectra, we trim the first 30 minutes and fit the white light curve with an exponential ramp (the amplitude and timescale both being free parameters), a linear trend, a linear function of the trace position in both directions, a transit model computed by \texttt{Batman} \citeApp{kreidberg_batman_2015} (with the transit depth and the second term of the quadratic limb darkening coefficients being free parameters), and an error inflation parameter.  We average the transit times, a/Rs, and b from the two white light curve fits to obtain fiducial transit parameters, which we adopt for all spectroscopic fits.  For the spectroscopic fits, we start by dividing the light curves by the systematics model found for the white light curve.  We then fit the light curves in exactly the same way as the white light curve, except that the transit parameters (transit times, a/Rs, and b) are fixed.  For the F322W2 data, we mask out the starspot crossing in all fits.  The starspot crossing is presumed to span phases -0.000519 to +0.00695, in accordance with the other analyses.

\subsubsection*{Analysis D}

This analysis uses the $Eureka!$ \citeApp{bell_eureka_2022} software package. The default JWST pipeline settings were adopted for stages 1 and 2. We omitted the jump step, which improved the quality of the resulting light curve for a bright target with few groups. In Stage 3, we first masked all pixels for which the 'DO$\_$NOT$\_$USE’ data quality flag was raised by the JWST pipeline. We trimmed the subarray to include pixels 4–1704 in the spectral direction for F322W2, and 780–2000 in the spectral direction for F444W. We then performed an outlier rejection along the time axis for each individual pixel in a segment using a 10$\sigma$ threshold. We corrected the 1/f noise in each of the four amplifier regions by subtracting the median flux in each row, excluding pixels containing flux from the star (columns 0 to 512 for F444 and 1850 to 2048 for F322W2). To extract the flux, we used optimal extraction, with an aperture centered on the brightest pixel in the spatial direction with a half-width of 7 pixels. For background subtraction, we exclude a region centered on the trace with a half-width of 11 pixels and use the remaining pixels to compute the background using a first-order polynomial for each column along the trace. We fit the spectroscopic light curves with models that included both the transit and the systematic noise. For these spectroscopic light curves, we summed the data into 15 nm bins (about 15 pixels). The parameters were estimated with a Markov chain Monte Carlo fit, using the \texttt{emcee} \citeApp{foreman-mackey_emcee_2013} Python package. We use the quadratic limb-darkening law, with the first parameter fixed to the value calculated using \texttt{ExoTic-ld} \citeApp{grant_exo-ticexotic-ld_2022} and the second fitted as a free parameter. We calculate our transit model using the package \texttt{Batman} \citeApp{kreidberg_batman_2015} assuming a circular orbit. We fix a/r* and inc to the best-fit value from the white light curve fits. We also include a systematic model in the form of a second-order polynomial in y-position and a linear + exponential trend in time. We drop the first 100 integrations because of the strong ramp and mask-out integrations 6500-7200 where a spot crossing occurs in our F322W2 data.

\subsubsection*{SW channel photometric observations}

We performed two independent analyses on the simultaneous short-wavelength (SW) channel photometric observations. Both F444W and F322W2 visits used the WLP4 weak lens and F212N2 narrow-band filter centered at 2.12 $\mu m$. The first visit used the NRCA1 detector and the second visit used the NRCA3 detector. Both analyses first applied row-by-row background subtraction to remove 1/f noise at the group level and then performed aperture photometry on the defocused images to produce the transit light curves. Both analyses found the light curves from the two visits to exhibit different temporal ramp behaviors likely due to the use of different detectors. The best-fit transit depths from the two SW visits also show offsets compared to the LW channel spectra. Since the SW observations are photometric, potential offsets can not be marginalized during the modeling. The F212N2 narrow-band filter also has very limited wavelength coverage and does not provide substantial additional information compared to the LW spectra. For those two reasons, both SW datasets were excluded from the analysis. \\

\section*{Atmospheric Modeling}

To interpret the transmission spectra of HD~189733b we use atmospheric radiative transfer models for a planet in transmission geometry coupled with a Bayesian parameter estimation framework in what is known in the field as ``atmospheric retrievals''. An atmospheric retrieval is composed of a parametric model (i.e., `` a forward model'') with a series of physical and chemical assumptions that try to capture the nature of the planetary atmosphere, and a statistical method to perform the parameter estimation such as Nested Sampling \citeApp{skilling_nested_2006} (see e.g., ref. \citeApp{madhusudhan_atmospheric_2018} for a review). This methodology results in posterior probability distributions for the model parameters capturing the most likely model parameters that explain the data, as well as the Bayesian evidence for the model considered ($\mathcal{Z}$). The Bayesian evidence, as well as other model comparison metrics (see e.g., ref. \citeApp{welbanks_application_2023}), can be used to compare two or more model configurations and estimate a preference for one over the other. The ratio in model Bayesian evidence is generally converted to a corresponding $\sigma$ level and used to quantify the model preference or detection for different chemical species in the field \citeApp{trotta_bayes_2008,benneke_how_2013,welbanks_aurora_2021}.

The atmospheric models in an atmospheric retrieval vary in their complexity and number of free parameters. Typically, these parametric models assume a one-dimensional atmospheric column with multiple chemical abundances, a pressure-temperature profile, and some treatment for the presence of clouds and hazes that may mute the spectral features of the atmosphere. Generally, with increasing complexity and physical consistency in these atmospheric models the number of parameters is reduced and so is the flexibility (see e.g., \citeApp{fortney_framework_2013} for a review). Conversely, atmospheric models with a large number of free parameters are thought of as a data-driven alternative but may be susceptible to strong degeneracies (e.g., Refs. \citeApp{benneke_how_2013,line_influence_2016,heng_theory_2017, welbanks_mass-metallicity_2019}) and can result in solutions with possibly nonphysical solutions (e.g., Refs. \citeApp{macdonald_why_2020, welbanks_atmospheric_2022}).

\subsubsection*{1D-RCPE model grids}

For transmission spectroscopy, reliable abundance constraints with fully parametric atmospheric retrievals require measurements of multiple broadband absorption features for different molecules \citeApp{benneke_atmospheric_2012}. Since our observations of HD~189733b cover a single absorption feature of CO$_2$ and partial absorption features of H$_2$O and CO, we decide to overcome these degeneracies by not fitting directly the molecular gas abundances and pressure-temperature profile but instead, predict them under the assumption of one-dimensional radiative-convective-photochemical-equilibrium (1D-RCPE) given an elemental composition and irradiation. First, the assumption of radiative-convective equilibrium (RCE) \citeApp{guillot_radiative_2010} is employed to calculate the pressure-temperature profile of the atmosphere given the planet's internal temperature, the incident stellar flux, and the opacity structure of the atmosphere. Within this framework, equilibrium is achieved when the net flux divergence in the vertical direction is zero ref \citeApp{marley_cool_2015}. Then, the photochemical-equilibrium (PE) refers to the computation of the ``disequilibrium state'' of atmospheric chemistry as a result of the chemical kinetics due to photochemistry and vertical mixing.

\subsubsection*{Integrating VULCAN photochemistry and vertical mixing}

We perform these 1D-RCPE computations by coupling the Sc-CHIMERA RCE solver \citeApp{piskorz_ground-_2018, mansfield_unique_2021, iyer_sphinx_2023} and the VULCAN kinetics solver \citeApp{tsai_comparative_2021,tsai_vulcan_2017, tsai_photochemically_2023}. This methodology has recently been adapted to explain observations of JWST and derive the atmospheric bulk properties of metallicity and carbon-to-oxygen ratio \citeApp{carter_jwst_2023}. The resulting vertical mixing ratios and pressure-temperature profile of the planetary atmosphere are computed for a given incident stellar flux (which results from re-scaling a specific irradiation temperature $T_{\rm irr}$ to the PHOENIX \citeApp{husser_new_2013} stellar flux spectrum at the top of the atmosphere), atmospheric metallicity $[M/H]$ (where the bracket notation indicates $\log_{10}$(M/H) relative to solar), and carbon-to-oxygen ratio, C/O (where the solar value is 0.46; \citeApp{lodders_44_2009}). In our calculations, the metallicity term scales the solar abundances of all elements heavier than H and He with Ref. \citeApp{lodders_44_2009} as the reference for a solar composition. The C/O ratio is adjusted for each scaled metallicity such that the sum of carbon and oxygen is preserved. The mixing ratios for the chemical species considered are initialized under the assumption of thermochemical equilibrium, using the NASA CEA2 Gibbs free energy minimization solver from Ref. \citeApp{gordon_computer_1994}, for a given pressure-temperature profile and elemental abundances. The opacity sources included in our calculations are H$_2$/He collision-induced absorption, H/e-/H- bound/free-free continuum, and the line opacities for H$_2$O, CO, CO$_2$, CH$_4$, NH$_3$, H$_2$S, PH$_3$, HCN, C$_2$H$_2$, OH, TiO, VO, SiO, FeH, CaH, MgH, CrH, ALH, Na, K, Fe, Mg, Ca, C, Si, Ti, O, Fe+, Mg+, Ti+, Ca+, C+.

When considering the effects of chemical disequilibrium, we use the converged RCE-thermochemical-equilibrium state to set the elemental abundances to initialize the VULCAN kinetics solver. We use the stellar spectrum for HD~189733 from \citeApp{moses_disequilibrium_2011}, and a power law eddy diffusion profile as described in Ref. \citeApp{tsai_photochemically_2023} scaled to approximate the profiles in Ref. \citeApp{moses_disequilibrium_2011} resulting in vertical diffusivities (Kzz) between 1$\times 10^{9}$ and 5$\times 10^{11}$ cm$^2$/s. Once converged, the resulting mixing ratios modified by the VULCAN computations are fixed in the 1D-RCTE to calculate the equilibrium conditions for all other species not in VULCAN such as Na, and K, and the resulting pressure-temperature profile. We perform these equilibrium and equilibrium-to-disequilibrium calculations in a grid of $T_{\rm irr}$ (1090--1300\,K in steps of 15\,K), [M/H] (-0.5--2.25 in steps of 0.125), and C/O (0.1--0.75 in steps of 0.05) for two internal temperatures in the extreme edges of theoretical prediction \citeApp{thorngren_intrinsic_2019}: 200~K and 500~K. The resulting grids are composed of 4830 models.

\subsubsection*{Grid retrieval}

The parameter estimation is performed by post-processing the 1D-RCPE atmospheric states through a line-by-line opacity sampling at a spectral resolution of R=100,000 using a radiative transfer scheme for transmission geometry, using as opacity sources H$_2$-H$_2$/He CIA\citeApp{karman_update_2019}, H$_2$O\citeApp{polyansky_exomol_2018}, CO\citeApp{gordon_computer_1994}, CO$_2$\citeApp{huang__reliable_2014}, Na \citeApp{allard_kh_2016}, K\citeApp{allard_kh_2016}, SO$_2$\citeApp{underwood_exomol_2016},  CH$_4$\citeApp{hargreaves_accurate_2020}, NH$_3$\citeApp{coles_exomol_2019}, HCN\citeApp{harris_improved_2006}, C$_2$H$_2$\citeApp{chubb_exomol_2020}, and H$_2$S\citeApp{azzam_exomol_2016}. The volume mixing ratio profiles, as well as the pressure-temperature profiles, are computed using a tri-linear interpolation using the scipy \citeApp{virtanen_scipy_2020} {\tt RegularGridInterpolator} for a given  $T_{\rm irr}$, [M/H], and C/O and are then used to calculate the transmission spectrum of the planet's atmosphere. Additionally, we post-process the transmission spectrum to include the presence of inhomogeneous clouds and hazes \citeApp{line_influence_2016} by using a linear superposition of a parametric power-law and gray cloud prescription composed of a vertically uniform grey cloud opacity ($\kappa_{\rm cld}$) and a deviation to a scattering hazes \citeApp{etangs_rayleigh_2008} and a cloud/haze-free atmosphere. Additionally, we include a scaling to the planetary radius at a reference pressure of 1~bar. Finally, we include a parameter for a possible offset for the NIRCam F322W2 observations relative to the NIRCam observations. In total, our atmospheric retrieval has a total of nine free parameters.

Our exploration of the atmospheric grid over the full eleven free parameter space resulted in unconstrained estimates for the $T_{\rm irr}$ parameter. To determine whether the resulting estimates of [M/H], and C/O are dependent on the $T_{\rm irr}$ parameter, we performed retrievals assuming full day-night redistribution and an irradiation temperature of $T_{\rm irr}=T_{\rm eq}\approx1192~K$. We find that the resulting estimates of [M/H] and C/O are consistent whether the irradiation temperature parameter is a free parameter or assumed to be the planet's equilibrium temperature. Therefore, our fiducial model is the ten-parameter model (Supplementary Information Table \ref{SI-grid_priors})) assuming an irradiation temperature equal to the planet's equilibrium temperature. The resulting 1$\sigma$ constraints are [M/H]=$0.51^{+0.06}_{-0.06}$ and  C/O=$0.12^{+0.02}_{-0.01}$. The constraints resulting from grid explorations have been shown to result, in some cases, in biased estimates if not interpolated appropriately \citeApp{fisher_how_2022}. Here we have interpolated our grid in their volume mixing ratios and pressure-temperature profiles instead of their spectra. Furthermore, we adopt as our resulting constraints the 1\% to 99\% confidence interval of each parameter. Therefore, our 1\% to 99\% constraints on the metallicity and C/O of HD~189733b are  [M/H]=0.47 to [M/H]=0.68 (i.e., 3--5$\times$ solar) and C/O=0.1 to 0.14, respectively (e.g., subsolar values). The presence of clouds and hazes remains unconstrained in our grid retrievals, while the offset between the NIRCam F322W2 observations and F444W4 observations is constrained to 20$\pm$6 ppm at the 1$\sigma$ confidence interval. The best-fit models VMR and TP profiles from the four grids are shown in Supplementary Information Figures \ref{fig:grid_vmr}.

To estimate the model preference or `detection significance' of the different gases in the atmosphere of HD~189733b we perform these retrievals after removing the spectroscopic contribution of each gas at a time. The Bayesian evidence of these nested retrievals is then compared to the Bayesian evidence of the fiducial model and converted to a $\sigma$-significance \citeApp{benneke_how_2013,welbanks_aurora_2021}. These model comparisons result in detections of H$_2$O at 13.4$\sigma$, CO$_2$ at $11.2\sigma$, H$_2$S at $4.5\sigma$, and CO at 5$\sigma$.

\begin{figure*}
\centering
\includegraphics[width=0.8\textwidth,keepaspectratio]{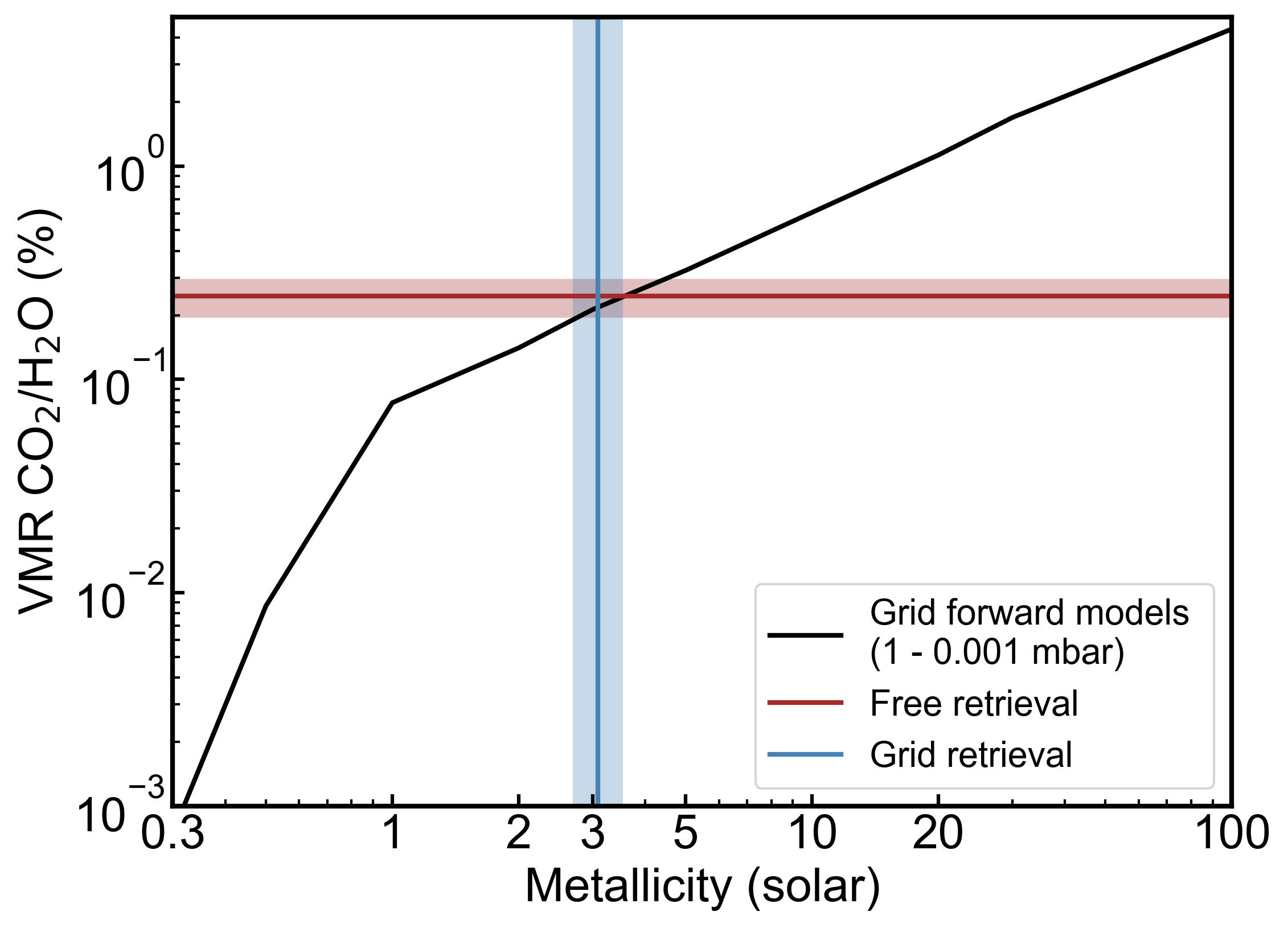}
  \caption{\textbf{The CO$_2$/H$_2$O ratio versus metallicity.} Transmission spectroscopy is directly shaped by the relative elemental abundance within the atmosphere from $\sim$1 to 0.001 mbar. Although free retrieval results can have degeneracies between different molecular abundances, they are robust at reflecting their relative ratios. The CO$_2$/H$_2$O from the free retrieval (brown) is consistent to within 1 sigma (shaded region) of CO$_2$/H$_2$O value at $\sim$3 time solar metallicity from the grid models. As we do not expect CO$_2$ or H$_2$O abundance to vary significantly from the equilibrium chemistry predictions for this planet, this agreement shows that free retrieval is consistent with the super-stellar metallicity inferred by the grid retrieval.}
  \label{fig:vmr_ratio}
\end{figure*}

\subsubsection*{Free retrieval}

We further explore the reliability of our detections and constraints by using the more flexible approach in which the volume mixing ratios are free parameters rather than determined by assumptions of chemical equilibrium as with the 1D-RCPE approach. Here we use the CHIMERA transmission retrieval tool \citeApp{line_systematic_2013, line_influence_2016, kreidberg_detection_2015, kreidberg_water_2018}. The transmission spectrum is generated using the same parallel-plane geometry as the 1D-RCPE post-processed models above, with the same prescription for inhomogeneous clouds and hazes. The difference here is that the volume mixing ratios for H$_2$O, CH$_4$, CO, CO$_2$, NH$_3$, HCN, H$_2$S, SO$_2$, and C$_2$H$_2$ are retrieved as free parameters assumed to be constant with height. Additionally, the pressure-temperature structure of the atmosphere is parameterized following the three-parameter version of the analytic \citeApp{parmentier_non-grey_2014} pressure-temperature profile (Supplementary Information Table \ref{SI-free_priors}).

The free retrieval on the NIRCAM F322W2 and F444W observations of HD~189733b results in constrained abundances of $\log_{10}(\text{H}_2\text{O})=-3.69^{+0.76}_{-0.46}$, $\log_{10}(\text{CO})=-4.13^{+0.88}_{-0.56}$, $\log_{10}(\text{CO}_2)=-6.24^{+0.67}_{-0.42}$, and  $\log_{10}(\text{H}_2\text{S})=-4.50^{+0.57}_{-0.35}$. The abundances of CH$_4$, NH$_3$, HCN, C$_2$H$_2$, and SO$_2$ are not constrained with 99\% upper limits in their logarithmic volume mixing ratios of -7.9, -5.3, -6.2, -6.3, and -7.0, respectively. The cloud and haze properties remain unconstrained. Finally, we retrieve an offset between the 322W2 and F444W observations of $-28.31^{+6.64}_{-7.15}$ppm.

\subsubsection*{Compare Grid to free retrieval results}

Both grid and free retrievals detected H$_2$O, CO$_2$, H$_2$S, and CO, but the free retrieval has lower absolute abundances and larger uncertainties (Figure \ref{fig:fig3}) due to its flexibility of allowing each molecule abundance to vary independently. The free retrieval can simultaneously increase or decrease elemental abundances for all molecules to achieve similar fits to the spectrum as shown in the free retrieval posterior (Figure \ref{fig:free_corner_inflate}). Since the metallicity is calculated by adding all elemental abundance and comparing them to that of solar values, this translates into $0.43^{+2.26}_{-0.28}$ times solar metallicity which is lower than $3.23^{+0.48}_{-0.42}$ times solar metallicity from the grid. However, this discrepancy can be resolved based on our knowledge of how CO$_2$ and H$_2$O are expected to follow equilibrium chemistry prediction as they are both robust again photochemistry \citeApp{moses_disequilibrium_2011}. As atmospheric metallicity increases, CO$_2$ abundance is expected to increase relative to H$_2$O as shown with the black line in Extended Data Figure \ref{fig:vmr_ratio}. The relative abundance of molecules within the atmosphere directly corresponds to their relative feature size within the transmission spectrum \citeApp{sing_observational_2018, heng_theory_2017}. The CO$_2$ to H$_2$O relative abundance is therefore robust from the free retrieval result and the CO$_2$ to H$_2$O ratio is within 1 sigma of predicted values from the forward models at $\sim$3 times solar metallicity. 

Although both methods detected H$_2$S, the grid retrieval shows less H$_2$S contribution than the free retrieval (Supplementary Information Figure \ref{fig:new_residual_grid}, \ref{fig:new_residual_free}). This is because the grid can only change the H$_2$S abundance by varying the bulk metallicity while the free retrieval can independently increase H$_2$S abundance. The larger H$_2$S contribution from the free retrieval indicates the need for more H$_2$S relative to the bulk metallicity in the grid. To demonstrate that, we ran another grid retrieval with an independent scaling parameter for the H$_2$S abundance (Extended Data Figure \ref{fig:grid_corner_inflate_scale}). The resulting S/H is $10.48^{+3.25}_{-2.4}$ times solar which is additionally enhanced relative to the bulk $\sim3$ times solar metallicity. The overall fit is improved (Supplementary Information Figure \ref{fig:new_residual_grid_H2S}) and the grid retrieval with H$_2$S scaling is preferred by 3 sigma over the one without.

\begin{figure*}
\centering
\includegraphics[width=\textwidth,keepaspectratio]{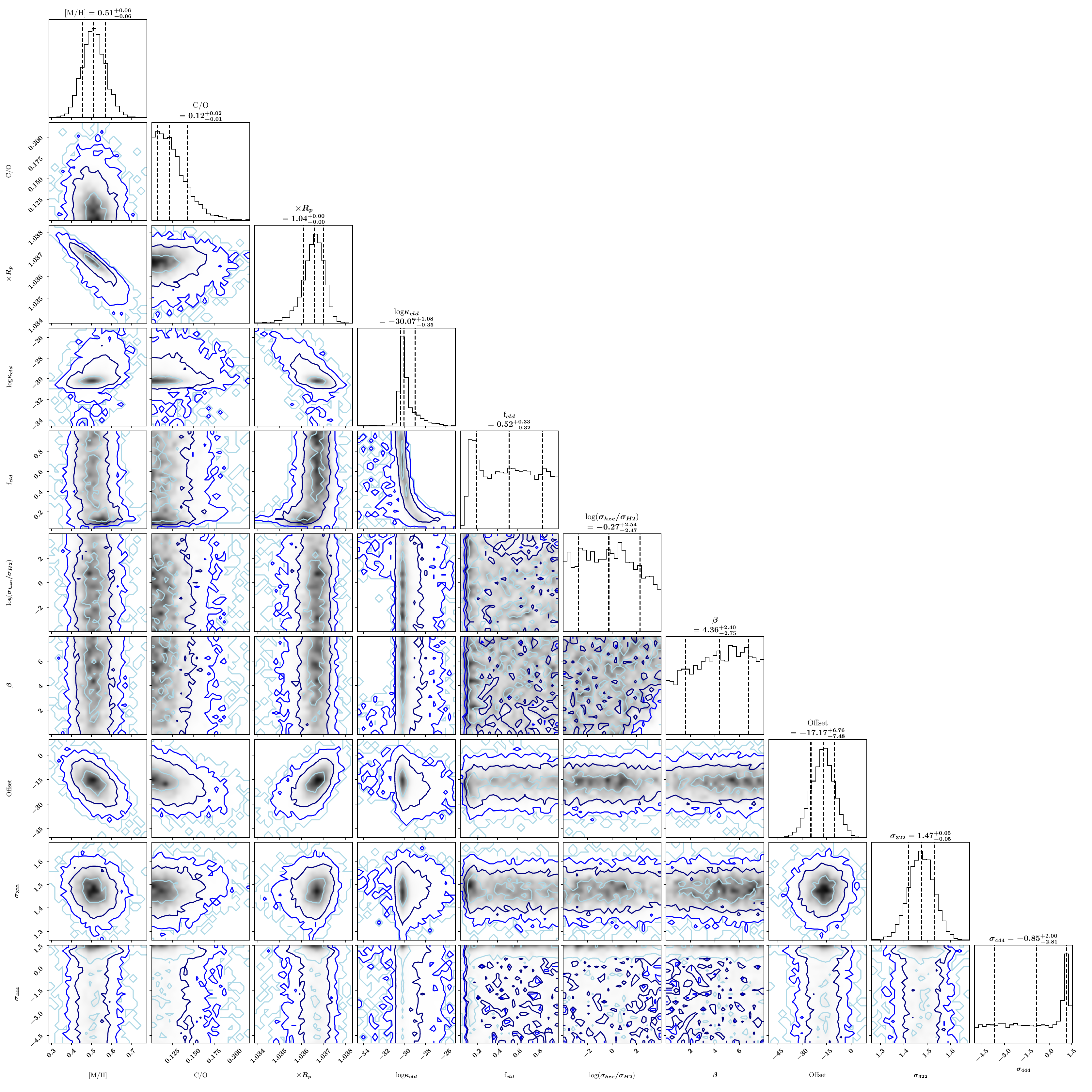}
  \caption{\textbf{Corner plot of the grid retrieval with error inflation.} Posterior distributions of the ten grid retrieval parameters.}
  \label{fig:grid_corner_inflate}
\end{figure*}

\begin{figure*}
\centering
\includegraphics[width=\textwidth,keepaspectratio]{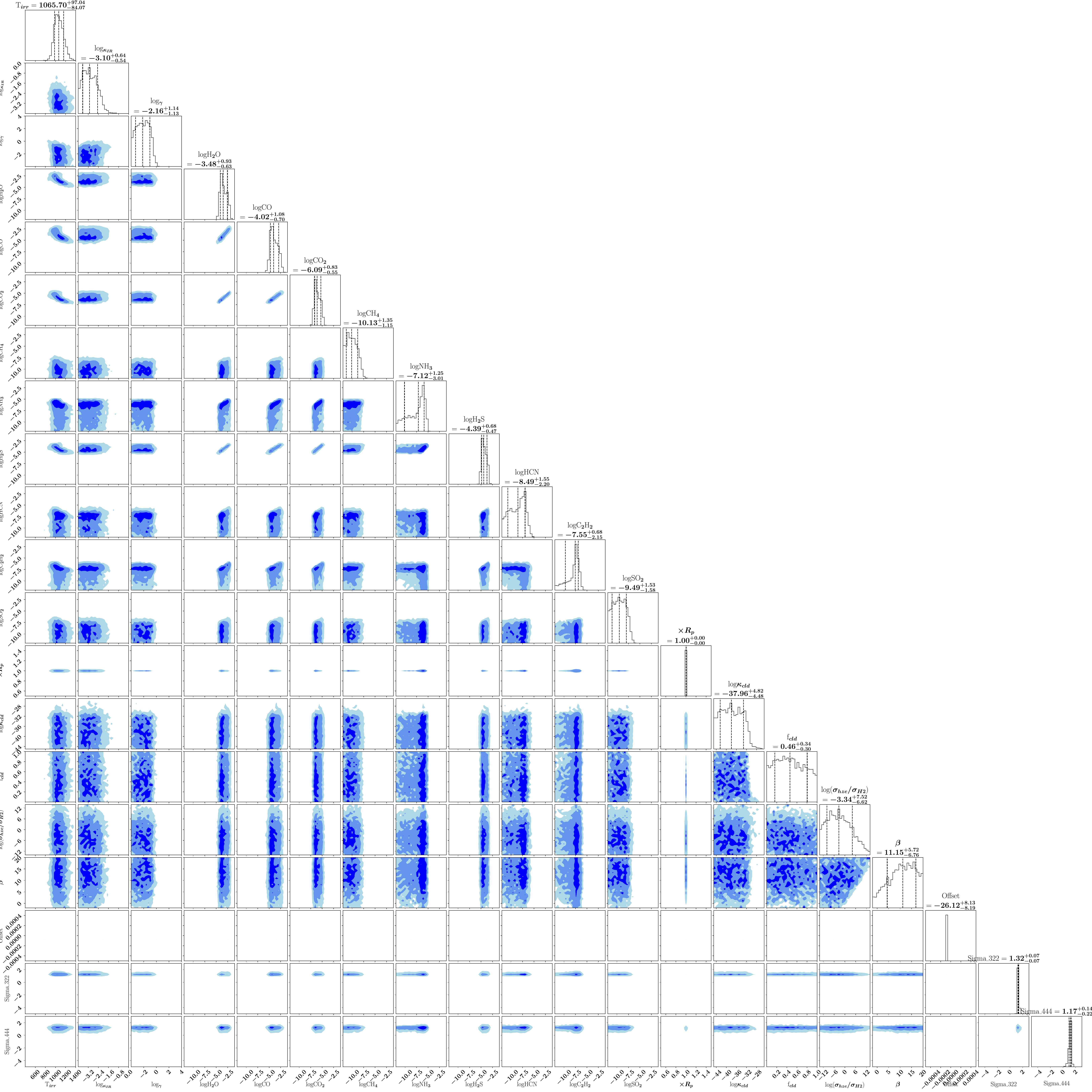}
  \caption{\textbf{Corner plot of the free retrieval with error inflation.} Posterior distributions of the twenty free retrieval parameters.}
  \label{fig:free_corner_inflate}
\end{figure*}

\begin{figure*}
\centering
\includegraphics[width=\textwidth,keepaspectratio]{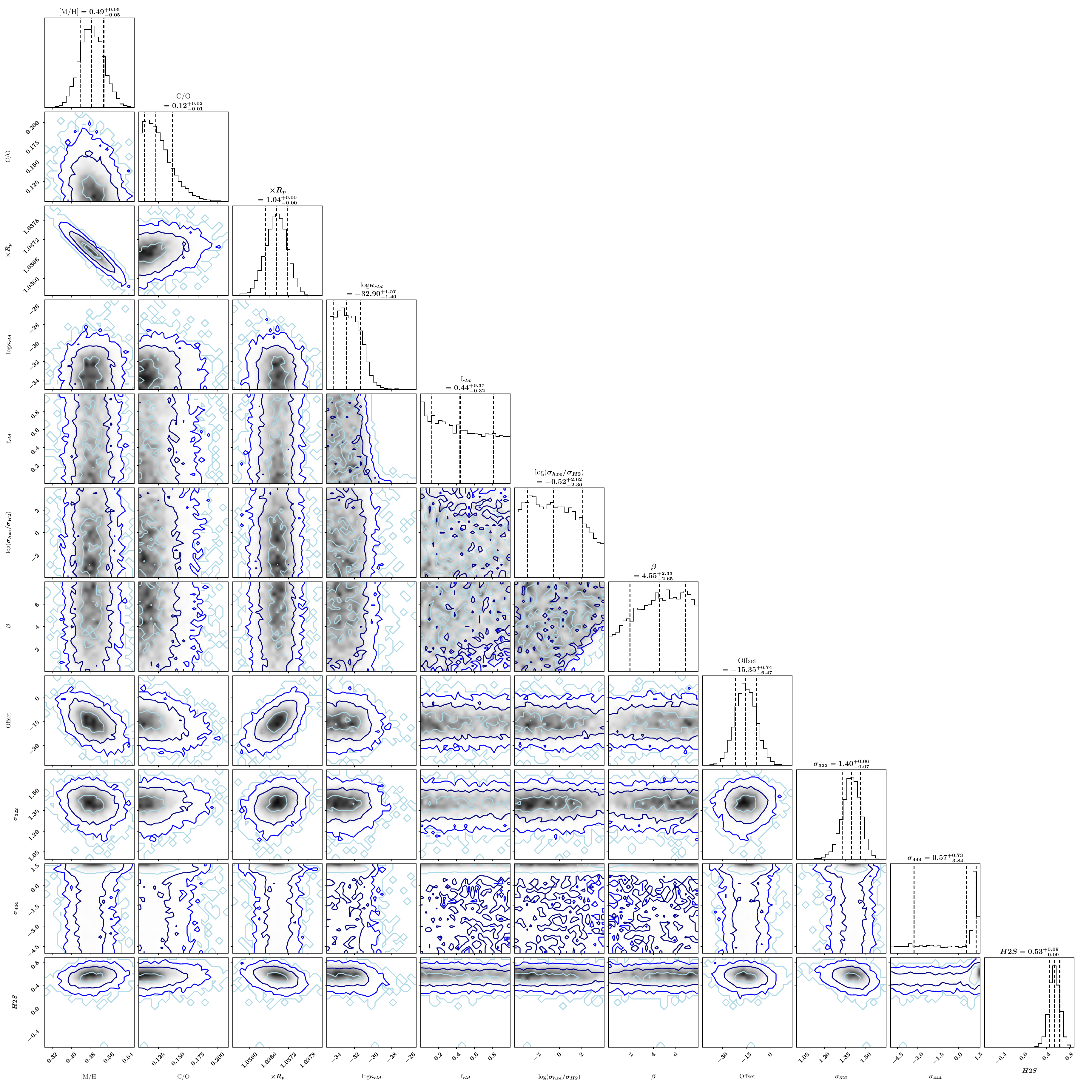}
  \caption{\textbf{Corner plot of the grid retrieval with H$_2$S scaling.} Same as Figure \ref{fig:grid_corner_inflate} but with an additional term of H$_2$S abundance scaling.}
  \label{fig:grid_corner_inflate_scale}
\end{figure*}

\section*{Long-term ground-based photometric monitoring}

We acquired 2091 photometric observations of HD~189733 from 2005 to 2022 with the T8, T10, and T15 0.80~m automatic photoelectric telescopes (APTs) at Fairborn Observatory in southern Arizona.  All three APTs are identically equipped with two-channel photometers that each use two EMI 9124QB bi-alkali photomultiplier tubes to measure stellar brightness simultaneously in the Str\"omgren $b$ and $y$ passbands.

The observations of HD~189733 were made differentially with respect to three nearby comparison stars.  We measure the difference in brightness between our program star HD~189733 (star~d) and the comparison stars (stars a, b, c) and create differential magnitudes in the following six combinations: d-a, d-b, d-c, c-a, c-b, and b-a.  Intercomparison of these six light curves shows that comp star~b (HD~191260) appears to be the most constant, so we present our results as differential magnitudes in the sense star~d minus star~b (d-b).

To improve the photometric precision of the individual nightly observations, we combined the differential $b$ and $y$ magnitudes into a single $(b+y)/2$ ``passband'', designated by the .by filename extension on the APT data files.  Our differential observations of HD~189733 are contained in the data file d-b.by.  The typical precision of a single nightly observation, as measured from pairs of constant comparison stars, typically ranges between 0.0010~mag and 0.0015~mag on good nights.  The T8 APT is described in \citeApp{henry_techniques_1999}, where further details of the telescope, precision photometer, and observing and data reduction procedures can be found.

Supplementary Information Figure \ref{fig:fig1_APT} plots the 2091 nightly d-b.by observations from 2005 to 2022 as small filled circles.  The mean of the nightly observations, 0.48762~mag, is plotted as the dashed line.  Since HD~189733 comes to opposition with the Sun around July~21, a significant portion of each yearly light curve is lost during the Summer Shutdown of Fairborn Observatory necessitated by the annual ``monsoon'' season in southern Arizona, which typically lasts from early July to early September.  Thus, each data section in Supplementary Information Figure \ref{fig:fig1_APT} represents only the first or second part of an observing season, beginning with 2005-B and ending with 2022-B.  The first 26 data sets (2005-B through 2018-A in Supplementary Information Table \ref{SI-APT}) were acquired with the T10 APT, the next 3 sets (2019-A through 2020-A) were acquired with T8, and the last 2 sets (2021-A and 2022-B) with T15.  The last few seasons are more sparsely covered due to significant downtime for necessary repairs and upgrades to the APTs. Supplementary Information Figure \ref{fig:fig1_APT} also plots the means of each data set with the large filled circles. The standard deviation of each mean magnitude is comparable to the size of the plot symbol.

Supplementary Information Table \ref{SI-APT} summarizes the 31 data sets.  The standard deviations of each set are given in column~5 and range from 0.00369 to 0.01224~mag, indicating night-to-night brightness variability in each set.  The mean magnitudes of each data set are given in column~6 and, as shown in Supplementary Information Figure \ref{fig:fig1_APT}, cover a range of 0.02082 mag, indicating significant year-to-year variability in HD~189733.  Period analysis of each data set using the method of \citeApp{vanicek_further_1971} confirms the presence of periodic variability in all but 2 data sets (column~7).  \citeApp{henry_nine_2022} show many examples of this method of period analysis in their work on $\gamma$~Doradus variable stars.  Least-squares sine fits on each set's best period and gives the peak-to-peak amplitudes listed in column~8.  These range from 0.0062 to 0.0310~mag.  The numbers in parentheses in columns 6-8 indicate the error bars on the last 2 digits.  The weighted mean period is $11.975\pm0.022$~days, not including the 2005-B period of 5.96~days.  During that time, HD~189733 apparently had active, spotted regions on opposite hemispheres, resulting in our period analysis finding half the true rotation period of the star.

Supplementary Information Figure \ref{fig:fig2_APT} presents a joint frequency analysis of all 31 d-b.by data sets. The 
data are normalized so that each data set has the same mean as the first. The top panel plots the normalized d-b.by observations.  The middle panel plots the frequency spectrum of those observations and finds the best period of $11.84497\pm0.00049$~days, which we take to be our best value for the rotation period of the star.  The bottom panel presents a phase curve of the d-b.by photometry on the best period with an arbitrary epoch equal to the HJD of the first photometric observation.  A least-squares sine fit on that period gives a peak-to-peak amplitude of $0.00699\pm0.00045$~mag, which can be considered an average of the rotational modulation over the course of our observations.

In this paper, we describe NIRCam spectroscopy of HD~189733 taken with the JWST on the nights of 25~August~2022 and 29~August~2022 at BJD median times of 2459816.9609 and 2459821.3984.  The epoch of these observations is shown in the top panel of Supplementary Information Figure \ref{fig:fig2_APT}, though the times of the two observations are not resolved at that scale.  We use the last two groups of our photometry from 2021-A and 2022-B, acquired with the T15 APT, to determine the rotational phases of the JWST observations.  These two data sets are shown in the top panel of Supplementary Information Figure \ref{fig:fig3_APT}.  The JWST observations occur between the two data sets.  The frequency spectrum of these data is shown in the middle panel of Supplementary Information Figure \ref{fig:fig3_APT}, which has a complex alias structure due to the gap in the observations. The best period for these two observing 
seasons is $11.83418\pm0.00812$~days, which is very close to the best period for the whole data set. The T15 observations phased to this period are shown in the bottom panel.  The phases of the two JWST observations are shown for comparison with the two arrows.  The photometric phases of the JWST observations computed against the time of minimum given in the bottom panel are 0.187 and 0.562, i.e., the first observation was acquired 2.21~days after photometric minimum and the second 6.65~days after minimum (0.73~days after maximum).

\subsubsection*{Data Availability}
The NIRCam data used in this paper are from JWST GO program 1633 (PI Deming) and are publicly available from the Mikulski Archive for Space Telescopes (MAST; \url{https://mast.stsci.edu}). Whitelight transit lightcurve, transit spectrum, and models are archived on Zenodo at \url{https://zenodo.org/records/11459715}.\\

\subsubsection*{Code Availability}
We used the following codes to reduce JWST NIRCam data: STScI JWST Calibration pipeline, \texttt{Eureka!}\citeApp{bell_eureka_2022}, numpy\citeApp{harris_array_2020}, scipy\citeApp{virtanen_scipy_2020}, and matplotlib\citeApp{hunter_matplotlib_2007}.\\


\bibliographystyleApp{sn-standardnature}
\bibliographyApp{references.bib}

\bmhead{Acknowledgments}
G.F. acknowledges support for this work provided by NASA through JWST GO program funding support. 

\bmhead{Author Contribution}
G.F. led the data analysis effort, contributed to the interpretation of the observations, and led the writing of the manuscript. L.W. led the modeling analysis effort including the grid and free retrievals using 1D-RCPE models. D.D. led the JWST GO 1633 program proposal and contributed to the data analysis effort. J.Inglis., M.Z., and E.S. contributed to the data analysis effort by providing additional data reductions for both NIRCam F322W2 and F444W wavelength channels. J.L., J.Ih., and M.N. performed 1D forward models and retrievals. J.M. performed photochemistry calculations. D.S. helped with creating the figures and text in the manuscript. M.L. and E.K. contributed to the model interpretation efforts. H.K., T.G., A.S., and D.L., are part of the proposal team and provided useful feedback for the project and the manuscript. G.H. provided the ground-based photometric monitoring data.

\noindent \textbf{Competing Interests Statement} The authors declare no competing interests.\\

\noindent\textbf{Additional information}\newline
\textbf{Correspondence and requests for materials} should be addressed to \href{Guangwei Fu}{mailto:guangweifu@gmail.com}.
\newline
\textbf{Reprints and permissions information} is available at \url{www.nature.com/reprints}.

\section*{Supplementary Information Figures}

\begin{figure*}[!htbp]
\centering
  \includegraphics[width=\textwidth,keepaspectratio]{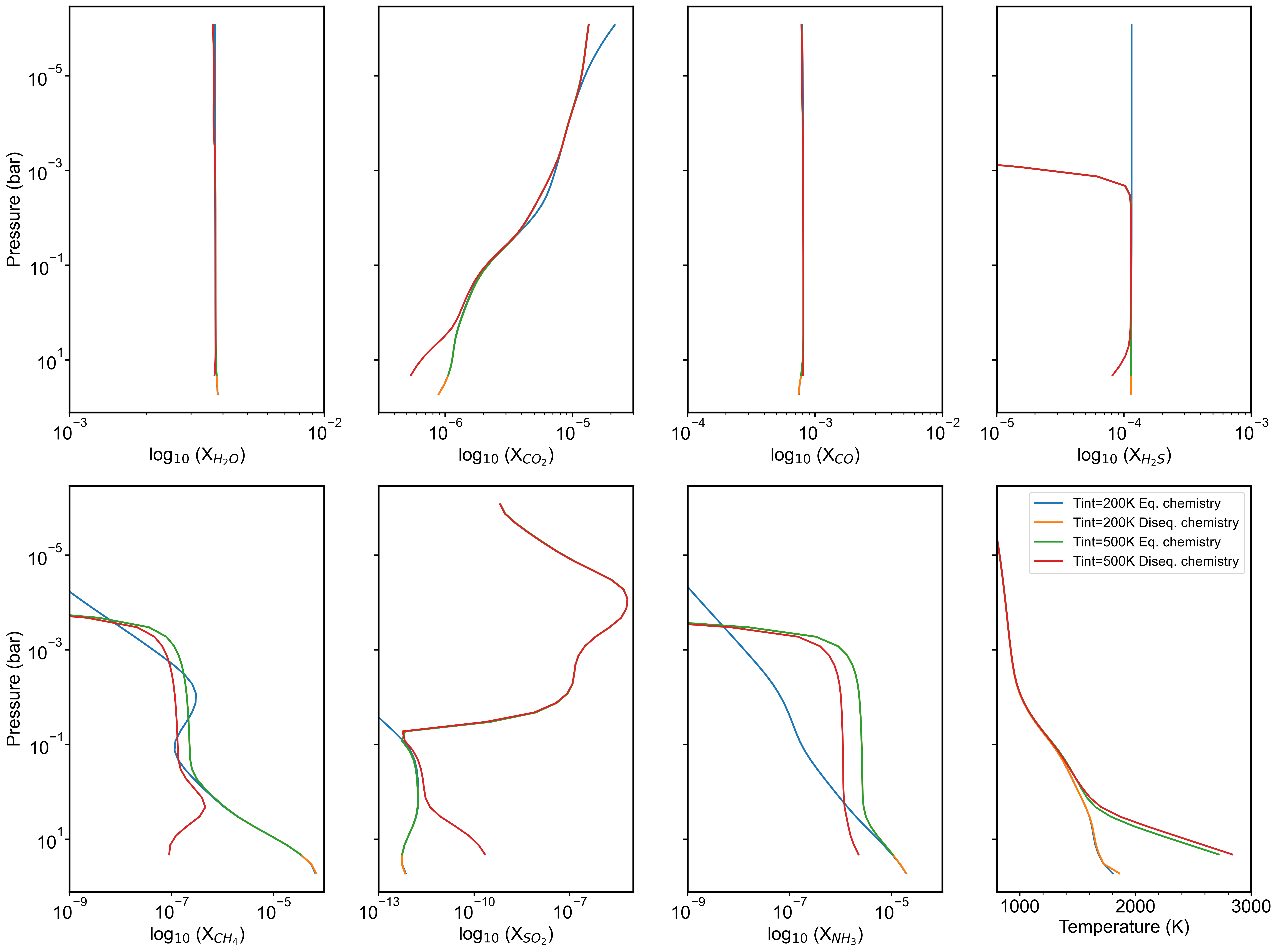}
  \caption{\textbf{The best-fit models VMR from the four 1D-RCPE grids.} We generated four 1D-RCPE grids with Tint = 200K and 500K under chemical equilibrium and disequilibrium assumptions. The four VMR and TP profiles in each panel share the same heat redistribution factor (0.958), metallicity (logZ=0.625), and C/O (0.15). The two disequilibrium chemistry models share the same Kzz vertical mixing profiles. The pressure levels probed by the transmission spectrum range from 10$^{-2}$ to 10$^{-4}$ bars.}
  \label{fig:grid_vmr}
\end{figure*}

\begin{figure*}
\centering
  \includegraphics[width=\textwidth,keepaspectratio]{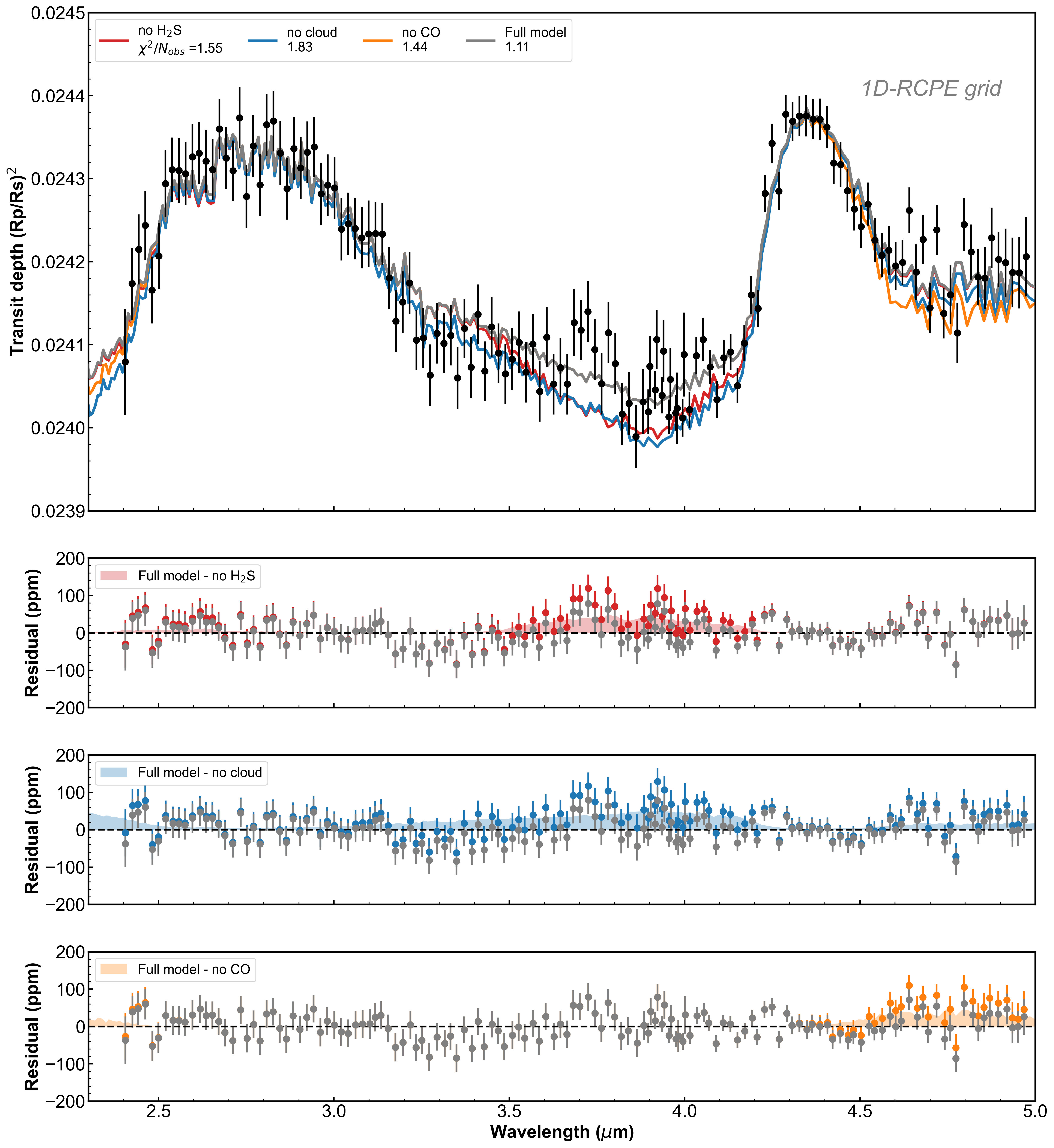}
  \caption{\textbf{Best fit and residuals of the grid retrieval.} The top panel shows the best fit full model and models with H$2$S, cloud, and CO subtracted. The bottom three panels show their respective residuals relative to the full model residual. The shaded color regions correspond to the wavelength region where each opacity source contributes.}
  \label{fig:new_residual_grid}
\end{figure*}

\begin{figure*}
\centering
  \includegraphics[width=0.9\textwidth,keepaspectratio]{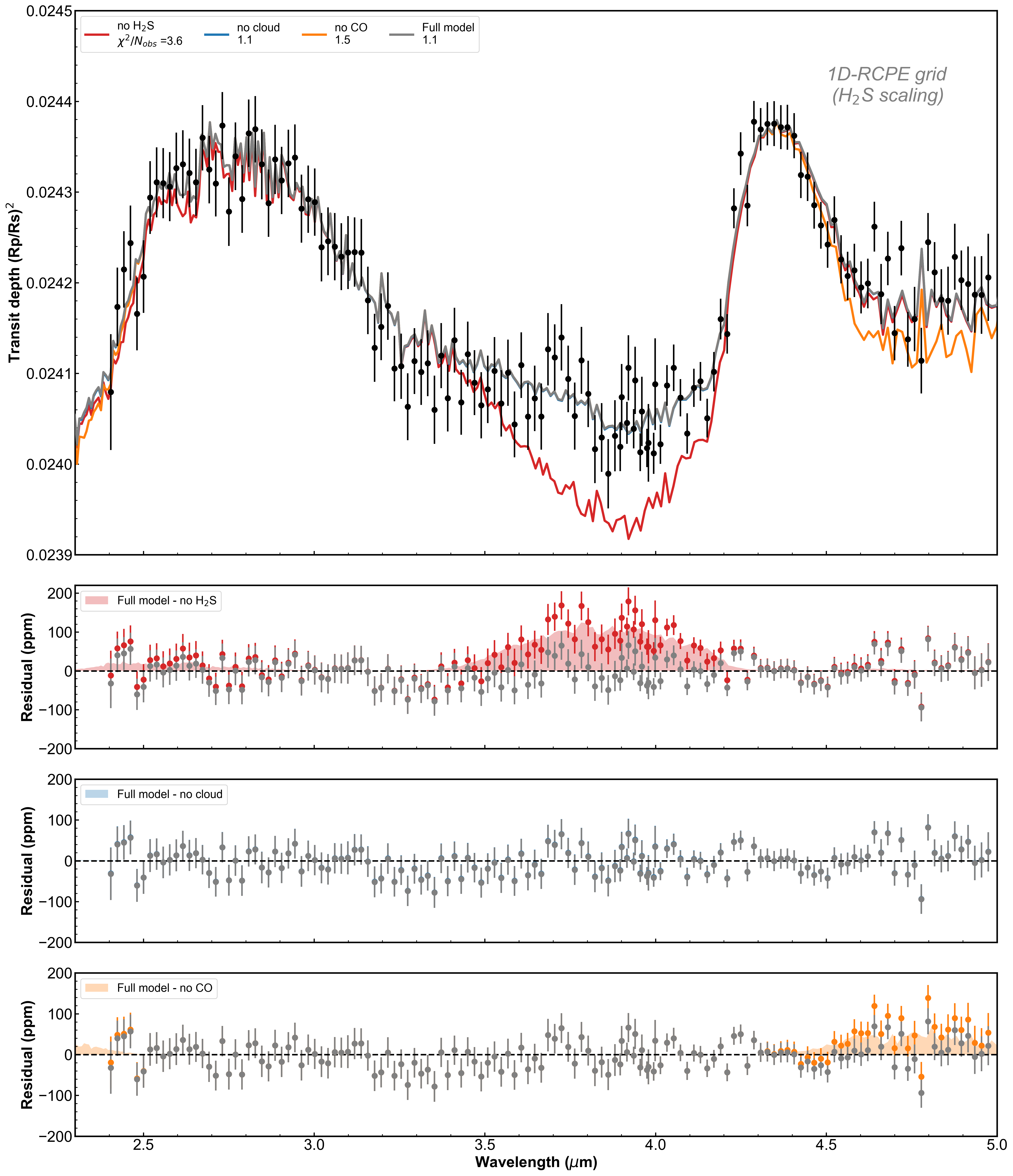}
  \caption{\textbf{Best fit and residuals of the grid retrieval with H$_2$S scaling.} The same as figure \ref{fig:new_residual_grid} but with additional free H$_2$S scaling parameter. The higher contribution of H$_2$S between 3.5 to 4.1 $\mu$m and 2.4 to 2.7 $\mu$m is evidence that enhanced H$_2$S abundance is needed and H$_2$S is preferred over clouds by the data.}
  \label{fig:new_residual_grid_H2S}
\end{figure*}

\begin{figure*}
\centering
  \includegraphics[width=\textwidth,keepaspectratio]{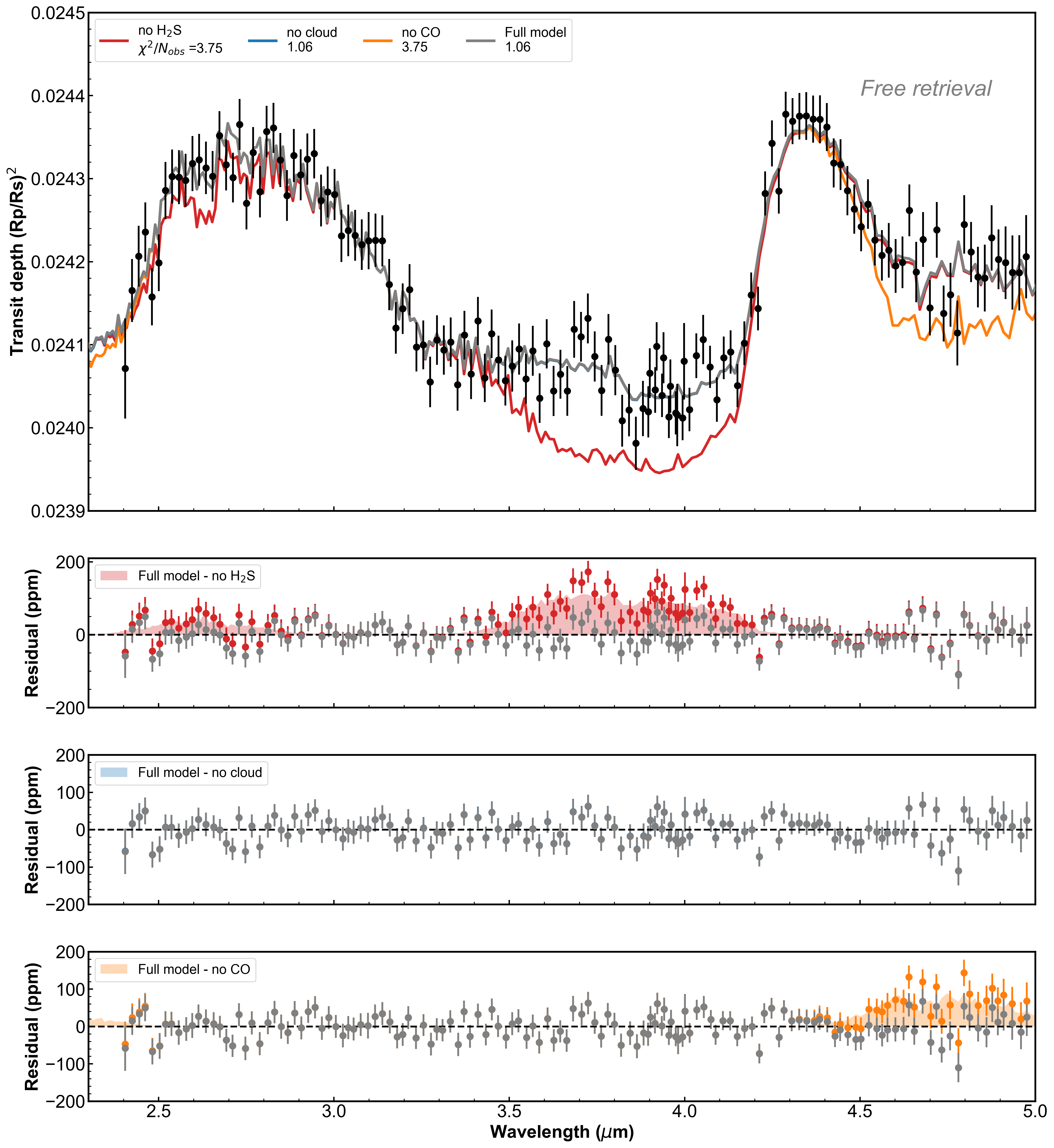}
  \caption{\textbf{Best fit and residuals of the free retrieval.} The same as figure \ref{fig:new_residual_grid} but for the free retrieval.}
  \label{fig:new_residual_free}
\end{figure*}

\begin{figure}
\centering
  \includegraphics[width=0.8\textwidth,keepaspectratio]{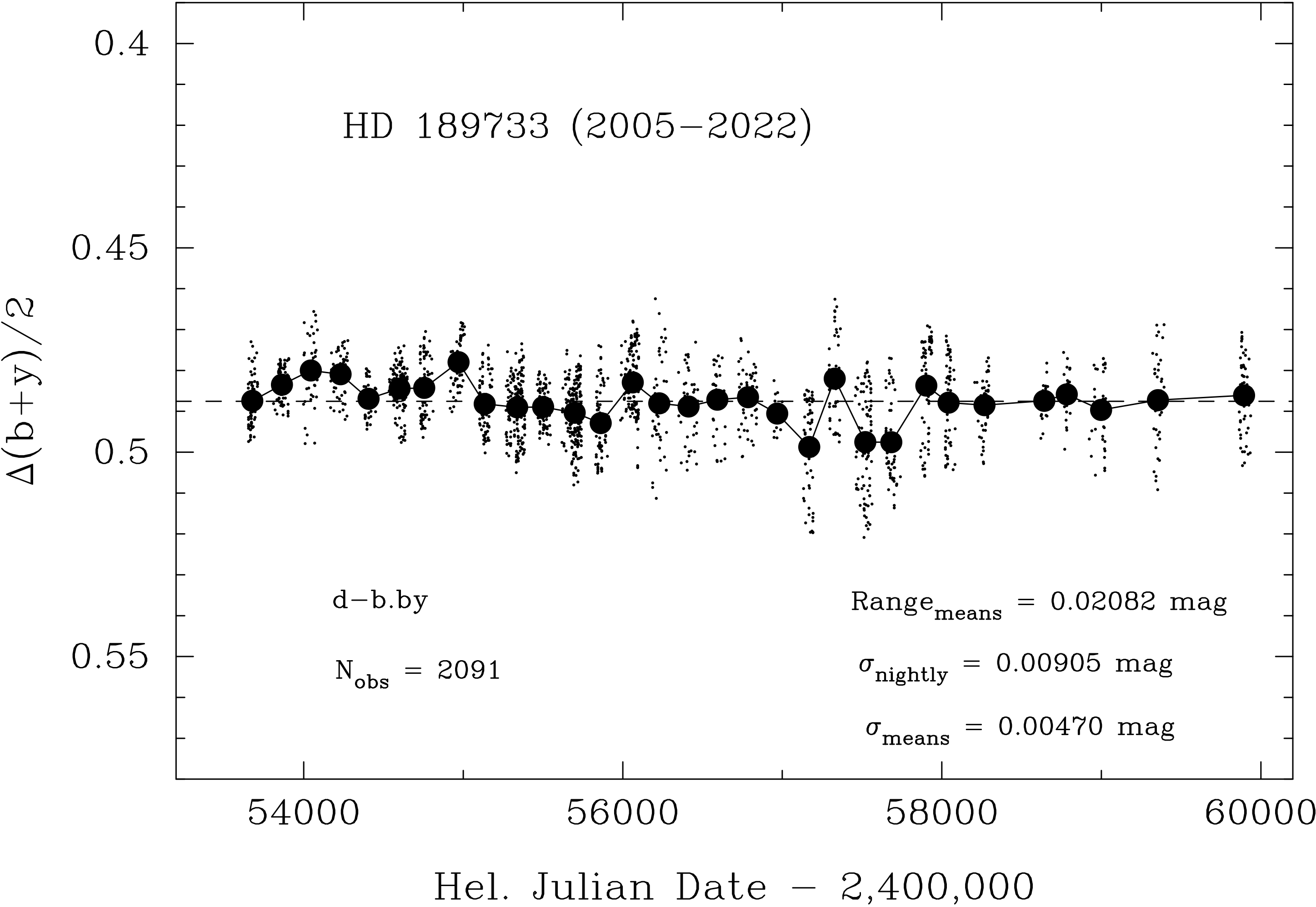}
  \caption{\textbf{Long-term photometric monitoring of HD 189733 from 2005 to 2022.} The nightly Str\"omgren $(b+y)/2$ band photometry (d-b.by) of HD~189733 (small filled circles) scatter about their mean (dashed line) with a standard deviation of 0.00905~mag.  Seasonal means from the 31 data sets (large filled circles) scatter about their mean with a standard deviation of 0.00470~mag.  Together, these results indicate low-amplitude variability on both night-to-night and year-to-year timescales.}
  \label{fig:fig1_APT}
\end{figure}

\begin{figure}
\centering
  \includegraphics[width=0.7\textwidth,keepaspectratio]{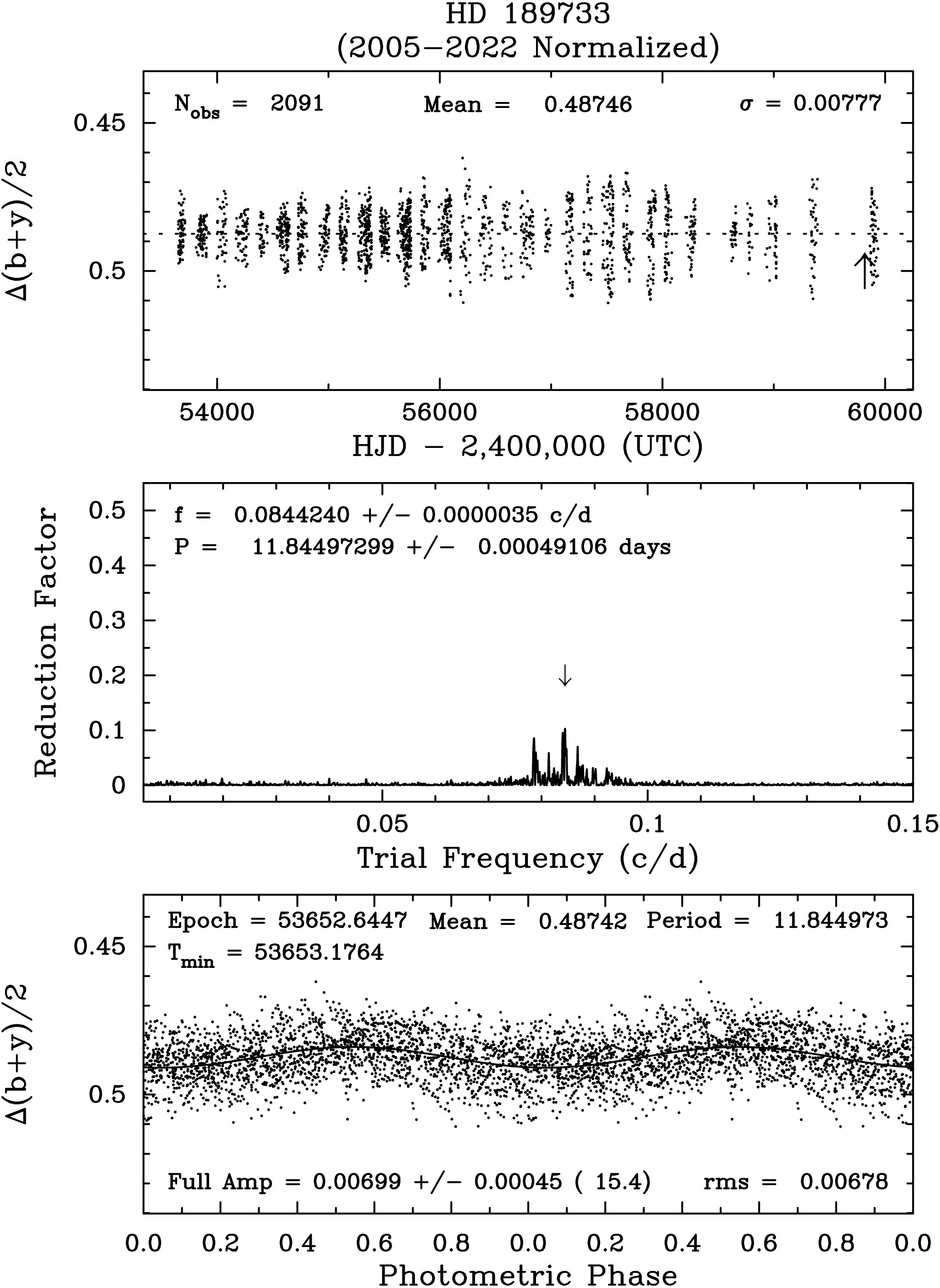}
  \caption{\textbf{HD 189733 brightness modulation period analysis.} Period analysis of all 31 normalized d-b.by data sets for HD~189733.  The top panel plots the normalized data, and the arrow indicates the epoch of our JWST observations.  The middle panel plots the frequency spectrum and finds a best period of $11.84497\pm0.00049$~days, which we take to be our best value for the rotation period of the star.  The bottom panel presents a least-squares sine fit on the normalized data phased with the best period of 11.84497~days.}
  \label{fig:fig2_APT}
\end{figure}

\begin{figure}
\centering
  \includegraphics[width=0.7\textwidth,keepaspectratio]{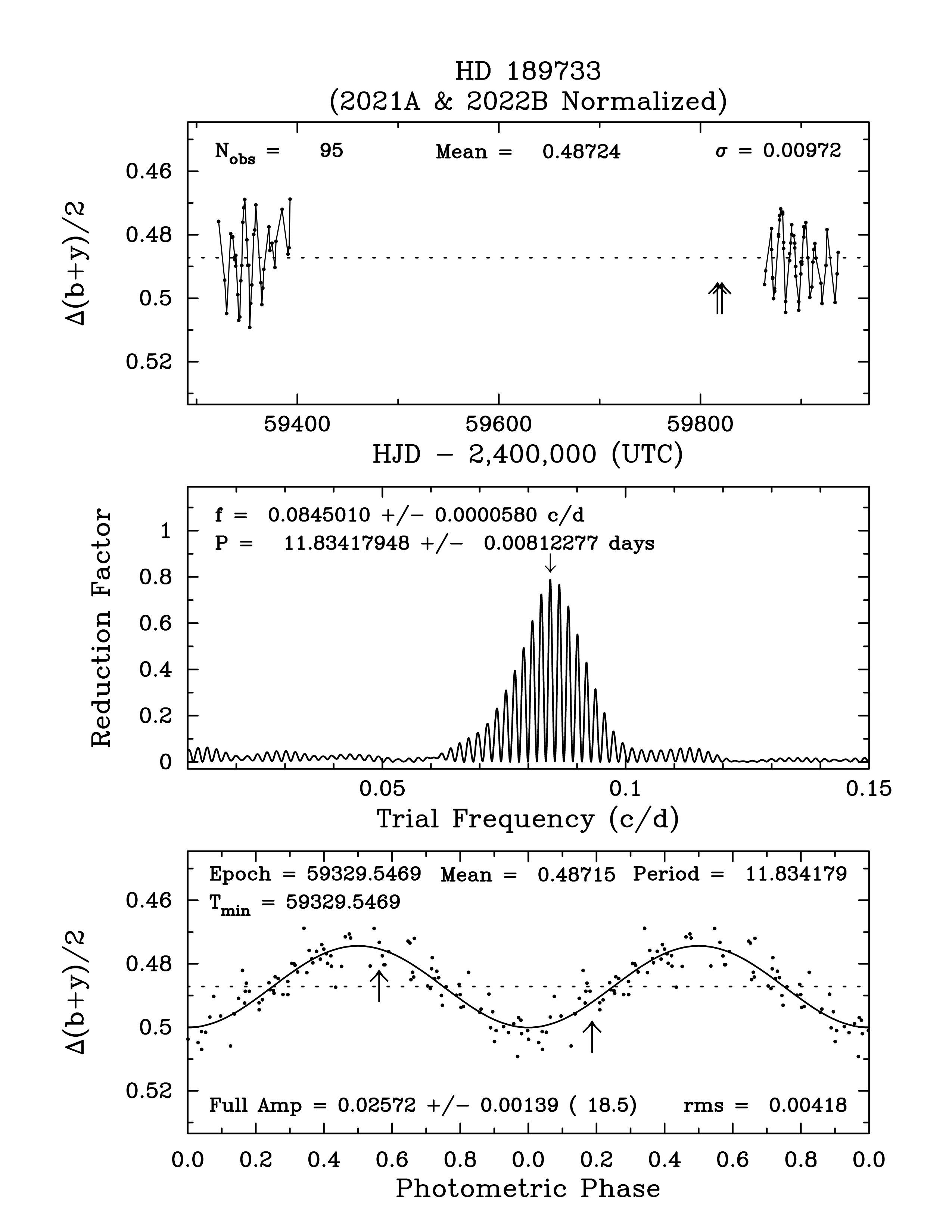}
  \caption{\textbf{HD 189733 photometric monitoring near JWST observations.} Period analysis of the T15 observations from 2021-A and 2022-B. The normalized observations are shown in the top panel along with the times of the JWST observations.  The frequency spectrum in the second panel gives a best period for these two data sets of 11.83418~days.  The arrows in the phase curve plotted in the bottom panel show the times of the JWST observations relative to the photometric time of minimum. The left arrow is for visit 2 with F322W2 and the right arrow is for visit 1 with F444W.}
  \label{fig:fig3_APT}
\end{figure}

\begin{table}[]
\begin{tabular}{cccc}
\hline
\textbf{Parameter} & \textbf{Lower Bound} & \textbf{Upper Bound} & \textbf{Prior} \\ \hline
T$_{irr}$ (K)	&	400	&	1400	&	uniform	\\
log$K_{IR}$	&	-4	&	0	&	log-uniform	\\
log$_\gamma$	&	-4	&	4	&	log-uniform	\\
logH2O	&	-12	&	-1	&	log-uniform	\\
logCO	&	-12	&	-1	&	log-uniform	\\
logCO2	&	-12	&	-1	&	log-uniform	\\
logCH4	&	-12	&	-1	&	log-uniform	\\
logNH3	&	-12	&	-1	&	log-uniform	\\
logH2S	&	-12	&	-1	&	log-uniform	\\
logHCN	&	-12	&	-1	&	log-uniform	\\
logC2H2	&	-12	&	-1	&	log-uniform	\\
logSO2	&	-12	&	-1	&	log-uniform	\\
$\times$Rp	&	0.5	&	1.5	&	uniform	\\
log$_{10}$K$_{cld}$	&	-45	&	-25	&	log-uniform	\\
f$_{\text{cld}}$	&	0	&	1	&	uniform	\\
log($\sigma_{hze}/\sigma_{H2}$)	&	-14	&	14	&	log-uniform	\\
$\beta$	&	-2	&	20	&	uniform	\\
Offset (ppm)	&	-500	&	500	&	uniform	\\
log$\sigma_{322}$ (ppm)	&	-5	&	3	&	log-uniform	\\
log$\sigma_{444}$ (ppm)	&	-5	&	3	&	log-uniform	\\
\hline
\end{tabular}
\caption{Priors used for the free retrieval. log$_\gamma$, log$K_{IR}$ and $\beta$ are parameters used to describe the TP profile. log$_{10}$K$_{cld}$, f$_{\text{cld}}$ and log($\sigma_{hze}/\sigma_{H2}$) are for cloud parameterization. log$\sigma_{322}$ and log$\sigma_{444}$ are error inflation terms for each spectral mode respectively.}
\label{SI-free_priors}
\end{table}

\begin{table}[]
\begin{tabular}{cccc}
\hline
\textbf{Parameter} & \textbf{Lower Bound} & \textbf{Upper Bound} & \textbf{Prior} \\ \hline
[M/H]	&	-0.5	&	2.25	&	log-uniform	\\
C/O	&	0.1	&	0.75	&	uniform	\\
$\times$Rp	&	0.5	&	1.5	&	uniform	\\
log$_{10}$K$_{cld}$	&	-35	&	-25	&	log-uniform	\\
f$_{\text{cld}}$	&	0	&	1	&	uniform	\\
log($\sigma_{hze}/\sigma_{H2}$)	&	-4	&	4	&	log-uniform	\\
$\beta$	&	0	&	8	&	uniform	\\
Offset (ppm)	&	-500	&	500	&	uniform	\\
log$\sigma_{322}$ (ppm)	&	-5	&	3	&	log-uniform	\\
log$\sigma_{444}$ (ppm)	&	-5	&	3	&	log-uniform	\\
\hline
\end{tabular}
\caption{Priors used for the grid retrieval.}
\label{SI-grid_priors}
\end{table}

\begin{table}
\begin{tabular}{@{}cccccccc@{}}
\toprule
Data & & & Date Range  & Sigma & Mean & Period & Full Amplitude\\
Set & APT & $N_{obs}$ & (HJD $-$ 2,400,000)  & mag & mag & days & mag\\
\midrule
   2005-B  & T10 &  72 & 53652--53712 & 0.00581 & 0.48746(68)  &  5.96(04) & 0.0142(10) \\
   2006-A  & T10 &  90 & 53811--53922 & 0.00373 & 0.48348(39)  & 13.54(12) & 0.0062(09) \\
   2006-B  & T10 &  41 & 53998--54085 & 0.00831 & 0.48002(130) & 12.00(13) & 0.0215(18) \\
   2007-A  & T10 &  57 & 54169--54282 & 0.00504 & 0.48094(67)  & 11.44(09) & 0.0096(15) \\
   2007-B  & T10 &  38 & 54377--54452 & 0.00451 & 0.48692(73)  & 10.91(24) & 0.0116(12) \\
   2008-A  & T10 & 124 & 54536--54648 & 0.00516 & 0.48445(46)  & 11.73(08) & 0.0109(10) \\
   2008-B  & T10 &  80 & 54728--54810 & 0.00698 & 0.48429(78)  & 11.88(14) & 0.0164(12) \\
   2009-A  & T10 &  64 & 54917--55006 & 0.00574 & 0.47799(72)  & 12.14(13) & 0.0108(15) \\
   2009-B  & T10 &  86 & 55102--55181 & 0.00616 & 0.48817(66)  & 12.02(24) & 0.0080(16) \\
   2010-A  & T10 & 168 & 55268--55385 & 0.00709 & 0.48903(55)  & 12.33(06) & 0.0114(13) \\
   2010-B  & T10 &  87 & 55463--55545 & 0.00444 & 0.48897(48)  & 11.43(10) & 0.0063(12) \\
   2011-A  & T10 & 196 & 55618--55740 & 0.00709 & 0.49036(51)  & 12.13(06) & 0.0149(10) \\
   2011-B  & T10 &  69 & 55830--55911 & 0.00841 & 0.49288(101) & 11.76(11) & 0.0241(13) \\
   2012-A  & T10 & 127 & 55987--56100 & 0.00795 & 0.48291(71)  & 11.92(08) & 0.0184(11) \\
   2012-B  & T10 &  40 & 56185--56274 & 0.01198 & 0.48806(189) & 12.27(14) & 0.0296(35) \\
   2013-A  & T10 &  45 & 56350--56470 & 0.00822 & 0.48882(122) & 12.03(11) & 0.0180(26) \\
   2013-B  & T10 &  31 & 56565--56641 & 0.00845 & 0.48714(152) & 12.16(27) & 0.0185(28) \\
   2014-A  & T10 &  59 & 56725--56840 & 0.00642 & 0.48655(84)  & 11.77(09) & 0.0110(20) \\
   2014-B  & T10 &  19 & 56947--56998 & 0.00369 & 0.49056(85)  &    &     \\
   2015-A  & T10 &  58 & 57132--57194 & 0.01114 & 0.49872(146) & 11.29(14) & 0.0239(30) \\
   2015-B  & T10 &  43 & 57293--57364 & 0.01010 & 0.48201(154) & 12.13(23) & 0.0245(18) \\
   2016-A  & T10 &  85 & 57459--57561 & 0.01224 & 0.49753(133) & 11.97(11) & 0.0319(13) \\
   2016-B  & T10 &  53 & 57649--57739 & 0.00983 & 0.49755(135) & 12.09(15) & 0.0256(20) \\
   2017-A  & T10 &  67 & 57863--57936 & 0.01088 & 0.48372(133) & 12.25(15) & 0.0287(12) \\
   2017-B  & T10 &  57 & 58021--58084 & 0.01017 & 0.48789(135) & 11.54(17) & 0.0268(16) \\
   2018-A  & T10 &  47 & 58207--58293 & 0.00670 & 0.48848(98)  & 12.13(20) & 0.0156(15) \\
   2019-A  & T8  &  31 & 58612--58662 & 0.00382 & 0.48741(69)  &    &    \\
   2019-B  & T8  &  20 & 58762--58804 & 0.00624 & 0.48585(140) & 12.50(53) & 0.0131(25) \\
   2020-A  & T8  &  42 & 58925--59026 & 0.00805 & 0.48966(124) & 11.79(12) & 0.0178(23) \\
   2021-A  & T15 &  40 & 59321--59392 & 0.01074 & 0.48724(170) & 11.66(17) & 0.0273(25) \\
   2022-B  & T15 &  55 & 59863--59936 & 0.00891 & 0.48608(120) & 11.85(15) & 0.0255(14) \\
\hline
\end{tabular}
\caption{Summary of APT photometric observations for HD 189733}
\label{SI-APT}
\end{table}

\end{document}